\documentclass[usenatbib,usegraphicx]{mn2e}
\usepackage{aas_macros}
\usepackage{rotating}
\usepackage{longtable}
\usepackage{times} 

\def\lsim{~\raise0.3ex\hbox{$<$}\kern-0.75em{\lower0.65ex\hbox{$\sim$}}~}
\def\gsim{~\raise0.3ex\hbox{$>$}\kern-0.75em{\lower0.65ex\hbox{$\sim$}}~}

\newcommand{\kms}{\,km\,s$^{-1}$~}
\newcommand{\kmseol}{\,km\,s$^{-1}$}

\def\gs{\mathrel{\raise0.35ex\hbox{$\scriptstyle >$}\kern-0.6em \lower0.40ex\hbox{{$\scriptstyle \sim$}}}}
\def\ls{\mathrel{\raise0.35ex\hbox{$\scriptstyle <$}\kern-0.6em \lower0.40ex\hbox{{$\scriptstyle \sim$}}}}
\newcommand{\arcsecs}{\mbox{$^{\prime\prime}$}}
\newcommand{\parcsec}{\mbox{$\stackrel{\prime\prime}{\textstyle .}$}}

\newcommand{\Msolar}{\mbox{$M_{\odot}\,$}}
\newcommand{\Lsolar}{\mbox{$L_{\odot}\,$}}
\newcommand{\degs}{\mbox{$^{o}$}}              

\newcommand{\COJ}[2]{\mbox{CO({#1}$-${#2})}}
\newcommand{\COJJ}[2]{\mbox{CO$\,J={#1}\rightarrow {#2}$}}
\newcommand{\J}[2]{\mbox{$J={#1} - {#2}$}}

\newcommand{\LOJ}[2]{\mbox{$L'_{\mbox{\tiny{CO}}({#1} - {#2})}$}}
\newcommand{\LCO}{\mbox{$L'_{\mbox{\tiny{CO}}}$}}
\newcommand{\LCOunits}{K\,km\,s$^{-1}$\,pc$^2$}
\newcommand{\XCO}{\mbox{$X_{\mbox{\tiny{CO}}}$}}

\begin{document}

\title[An interferometric CO survey of luminous submm galaxies]
      {An interferometric CO survey of luminous submm galaxies}

\author[Greve et al.]{
\parbox[t]{\textwidth}{
\vspace{-1.0cm}
T.\ R.\ Greve,$^{\! 1,2}$ 
F.\ Bertoldi,$^{\! 3}$ 
Ian Smail,$^{\! 4}$ 
R.\ Neri,$^{\! 5}$
S.\ C.\ Chapman,$^{\! 2}$ 
A.\ W.\ Blain,$^{\! 2}$
R.\ J.\ Ivison,$^{\! 1,6}$ 
R.\ Genzel,$^{\! 7,8}$ 
A.\ Omont,$^{\! 9}$
P.\ Cox,$^{\! 10}$
L.\ Tacconi$^{7}$ \&
J.-P.\ Kneib$^{2,11,12}$
}
\vspace*{6pt}\\
$^1$ Institute for Astronomy, University of Edinburgh,
     Blackford Hill, Edinburgh EH9 3HJ, UK.\\
$^2$ California Institute of Technology,
     Pasadena, CA 91125, USA.\\
$^3$ Max-Planck Institut f\"{u}r Radioastronomie(MPIfR), Bonn, Germany.\\
$^4$ Institute for Computational Cosmology, University of Durham, South Road,
     Durham DH1 3LE, UK.\\
$^5$ Institut de Radio Astronomie Millim\'etrique (IRAM),
     St Martin d'H\`{e}res, France.\\
$^6$ Astronomy Technology Centre, Royal Observatory, Blackford Hill, 
     Edinburgh EH9 3HJ, UK.\\
$^7$ Max-Planck Institut f\"{u}r extraterrestrische Physik (MPE), 
     Garching, Germany.\\
$^8$ Department of Physics, University of California, Berkeley, USA.\\
$^9$ Institut d'Astrophysique de Paris, CNRS, Universit\'{e} de Paris, 
     Paris, France.\\	
$^{10}$ Institut d'Astrophysique Spatiale, Universit\'{e} de Paris Sud, 
     Orsay, France.\\
$^{11}$ Observatoire Midi-Pyr\'en\'ees, UMR5572,
  14 Avenue Edouard Belin, 31400 Toulouse, France.\\
$^{12}$Laboratoire d'Astrophysique de Marseille, UMR 6110, 
CNRS-Université de Provence,  Traverse du Siphon-Les trois Lucs, 
13012 Marseille, France.
\vspace*{-0.5cm}}

\date{\fbox{\sc Draft dated: \today\ }}
\date{Accepted ... ; Received ... ; in original form ...}

\pagerange{000--000}

\maketitle

\begin{abstract} 
In this paper we present results from an IRAM Plateau de Bure
millimetre-wave Interferometer (PdBI) survey for CO emission towards 
radio-detected submillimetre galaxies (SMGs) with known optical and
near-infrared spectroscopic redshifts.  Five sources in the redshift range $z\sim 1$--3.5 
were detected, nearly doubling the number of SMGs detected in CO.  We
summarise the properties of all 12 CO-detected SMGs, as well as 6 sources not detected in
CO by our survey, and use this sample to explore the bulk physical properties of the
SMG population as a whole.  The median CO line luminosity of the SMGs is $\langle \LCO \rangle
= (3.8 \pm 2.0)\times 10^{10}$\,\LCOunits. Using a CO-to-H$_2$ conversion
factor appropriate for starburst galaxies, this corresponds to a molecular gas
mass $\langle M(\mbox{H}_2)\rangle = (3.0\pm 1.6)\times 10^{10}\,\Msolar$ within a
$\sim 2$\,kpc radius, about four times
greater than the most luminous local ultraluminous infrared galaxies (ULIRGs)
but comparable to that of the most extreme high-redshift radio galaxies and QSOs. The median CO
{\sc fwhm} linewidth is broad, $\langle${\sc fwhm}$\rangle=780\pm 320$\,\kms,
and the SMGs often have double peaked line profiles, indicative of either a
merger or a disk. From their median gas reservoirs ($\sim 3\times 10^{10}\,\Msolar$) and
star-formation rates ($\gs 700\,\Msolar\,\mbox{yr}^{-1}$) we estimate a lower limit on the typical gas-depletion time scale
of $\gs 40$\,Myr in SMGs. This is marginally below the typical age expected for the starbursts
in SMGs, and suggests that negative feedback processes may play an 
important role in prolonging the gas consumption time scale. We find a statistically-significant correlation between the
far-infrared and CO luminosities of the SMGs which extends the observed correlation 
for local ULIRGs to higher luminosities and 
higher redshifts.  The non-linear nature of the correlation implies that SMGs
have higher far-infrared  to CO luminosity ratios, and possibly higher star-formation
efficiencies, than local ULIRGs. Assuming a typical CO source diameter of $\theta
\sim 0\parcsec5$ ($D\sim 4$\,kpc), we estimate a median dynamical mass of
$\langle M_{dyn}\rangle \simeq (1.2\pm 1.5)\times 10^{11}\,\Msolar$ for the SMG sample. Both the
total gas and stellar masses imply that SMGs are very massive systems,
dominated by baryons in their central regions. The baryonic and dynamical
properties of these systems mirror those of
local giant ellipticals and are consistent with numerical simulations
of the formation of the most massive galaxies.  We have been able to impose a
lower limit of $\gs 5 \times 10^{-6}$\,Mpc$^{-3}$ to the co-moving number
density of massive galaxies in the redshift
range $z\sim 2$--3.5, which is in agreement with results from recent spectroscopic surveys
and the most recent model predictions.
\end{abstract}

\begin{keywords}
   galaxies: starburst
-- galaxies: formation
-- cosmology: observations
-- cosmology: early Universe
\end{keywords}

\section{Introduction}
The discovery of extragalactic carbon monoxide (CO) rotational line emission
(Rickard et al.\ 1975) and, in particular, the first detections of CO at
cosmologically significant redshifts in IRAS\,F10214$+$4724 at $z=2.29$ 
(Brown \& Vanden Bout 1991; Solomon, Downes \& Radford 1992a), the Cloverleaf at $z=2.56$
(Barvainis et al.\ 1994) and BRI\,1202$-$0725 at $z=4.69$ (Omont et al.\ 1996) revealed 
the potential of CO as a tracer of molecular gas in the early Universe.  Since
those pioneering efforts, progress has been slow due to 
the severe observational obstacles: 
the faintness of the CO emission, except for  cases were the
source is gravitationally magnified; inaccurate spectroscopic redshifts coupled
with the small instantaneous bandwidth of most modern-day correlators; and in
some cases, an unfortunate combination of redshift and available receiver
coverage, as well as atmospheric transparency, at millimetre wavelengths.  As a result only 30-or-so
$z>1$ objects have been detected to date, most of
which have been  extremely luminous, often
gravitationally lensed, QSOs (e.g.\ Omont et al.\ 1996; Guilloteau et al.\ 1997, 1999; 
Downes et al.\ 1999; Cox et al.\ 2002; Bertoldi et al.\ 2003; Beelen et al.\ 2004) and high-redshift radio galaxies 
(HzRGs --- e.g.\ Papadopoulos et al.\ 2000; De Breuck et al.\ 2003a).

The slow increase in the number of CO detections 
contrasts with the rapid growth in samples of high-redshift galaxies
selected through continuum observations at
submillimetre (submm) wavelengths, which detect thermal emission from
dust. The advent of large-format submm/mm cameras,
SCUBA (Holland et al.\ 1999) and MAMBO (Kreysa et al.\ 1998), revealed the
presence of a significant population of dust-enshrouded, and therefore hitherto
undetected, galaxies at high redshifts (e.g.\ Smail, Ivison \& Blain 1997; 
Barger, Cowie \& Sanders 1999; Bertoldi et al.\ 2000). Today such observations are
routine, and several hundred submm-selected (or SCUBA) galaxies (SMGs)
have been discovered (e.g.\ Blain et al.\ 2002; Scott et al.\ 2002; Webb et al.\ 2003; Borys et
al.\ 2003).

Until recently, only two SMGs had been detected in CO (Frayer et al.\ 1998,
1999; Downes \& Solomon 2003; Genzel et al.\ 2003), largely owing to the extreme faintness of SMGs in the optical and the
difficulties in obtaining reliably spectroscopic redshifts.  However, a major
step forward was made by Chapman et al.\ (2003b, 2005) who used the
highly-efficient, blue-sensitive LRIS-B spectrograph on the Keck Telescope
to obtain spectroscopic redshifts for a large sample of SMGs.  This motivated a
major survey at the IRAM Plateau de Bure Interferometer (PdBI) to look for CO
emission from this high-redshift sample.  The observational
cycle involves: identifying a robust radio counterpart to an SMG in deep VLA
radio maps (see e.g.\ Ivison et al.\ 2002); placing a LRIS-B/Keck slit on this
position to obtain a spectroscopic redshift (Chapman et al.\ 2003b, 2005); frequently confirming the redshift in the
near-infrared, usually via redshifted H$\alpha$ (Simpson et al.\ 2004; Swinbank et al.\ 2004); and, finally, searching for
redshifted CO emission with PdBI.  The initial results from this CO programme
were described in Neri et al.\ (2003). Although an expensive process in terms of
telescope time, this is currently the only feasible and effective route for detecting CO from
SMGs.

CO observations can potentially provide unique information about the
enigmatic SMG population. First and foremost, CO traces the bulk of the molecular
gas in SMGs: the high-level ($J\ge 2$) transitions
arise in the warm and dense gas while lower level $J$ lines probe the quiescent and likely
cooler gas which may lurk in the outskirts of SMGs.  Detecting and mapping this
molecular emission enables us to precisely determine the spatial
and kinematic location of the gas-rich components within an SMG.
About two-thirds of SMGs are found to be large, morphologically complex systems
in the optical/near-infrared and/or the radio, with typically one or more companions  
(e.g.\ Smail et al.\ 1999; Ivison et al.\ 2002; Chapman et al.\ 2003c, 2004). 
For example, in one SMG, Neri et al.\ (2003) identified CO emission
coincident with a second, fainter radio source $\sim 4\arcsecs$ away from the
radio counterpart for which the spectroscopic redshift had been found.
Secondly, CO observations yield fairly accurate estimates of the amount of
molecular gas available to fuel the starburst and/or the AGN responsible for the
large far-infrared luminosities.  While the conversion factor between the CO
luminosity and molecular gas mass is uncertain, CO observations clearly
provide a much better constraint on the amount of gas present in SMGs than 
the estimates based on submm continuum observations and an adopted 
spectral energy distribution (SED) and gas-to-dust ratio.

The first two CO detections of SMGs revealed the presence of copious amount of
molecular gas ($\sim 10^{10}\,\Msolar$), suggesting that intense star formation
is occuring in these systems (Frayer et al.\ 1998, 1999).
Furthermore, from a reliable estimate of the gas reservoir in an SMG we can say
something about the gas exhaustion time-scale, i.e.\ the duration of
the submm-luminous phase.  This, in turn, allows us to make an educated
guess about the possible descendants of SMGs, and thus place them in an
evolutionary context with other high- and low-redshift galaxy populations.

Observations of the shape and width of CO lines also provides
important information about the kinematics in SMGs 
(Neri et al.\ 2003; Tacconi et al.\ 2005). In particular, if the CO
emission is spatially resolved we can use that, in conjunction with the width
of the line profile, to constrain the dynamical mass of the host
galaxy.  Estimates of the
dynamical mass based on CO are likely to be `cleaner' than estimates from
optical/near-infrared spectroscopy, which are prone to extinction by dust and
the effects of non-gravitational motions in the emitting gas, such as
outflows.  The best example of resolved CO emission in a
SMG (Genzel et al.\ 2003) shows gas extended on scales of $\sim3$--5\,kpc and that
most of the dynamical mass ($\sim 3\times 10^{11}\,\Msolar$) is baryonic.
Such large, widely-distributed gas reservoirs suggest that the brightest SMGs are not
merely high-redshift replicas of the local population of ULIRGs. However,
a representative picture of the gas distribution in SMGs will have to await high-resolution 
CO observations of a large sample of SMGs. 

Since CO observations provide a means of `weighing' galaxies at high redshifts,
both in terms of their baryonic gas mass content and their total
dynamical mass, they can be used to help piece together a picture of the mass
assembly of massive galaxies in the early Universe.  In the classical cold-dark
matter (CDM) scenario of structure formation (White \& Frenk 1991), massive
spheroidal galaxies are the end products of a gradual build-up of mass via
merging and, as a result, form late in the history of the Universe. However, as
pointed out by Genzel et al.\ (2003), at least some SMGs appear to be massive,
baryon-dominated galaxies already at $z\sim 2$--3.  The discovery of massive,
evolved early-type galaxies at $z\gs 1.5$ (e.g.\ Cimatti et al.\ 2004), which could be the descendants of $z\gs 2$
SMGs, suggests that some massive spheroids were not formed at $z\ls 1$. In
fact, if the brightest 25 per cent of the SMGs are massive baryonic systems,
then their abundance indicates that the build up of massive galaxies in the
early Universe was much faster than previously expected, and could pose a
serious challenge for current models of gas processing in galaxy formation (e.g.\ Cole et
al.\ 1994; Kauffmann et al.\ 1999; Baugh et al.\ 2004).

In this paper we present the most recent results from a systematic survey of CO
emission towards radio-identified SMGs with spectroscopic redshifts in the
range $z\sim 1$--3.5 (corresponding to 40--10\% of the age of
the Universe at the respective epochs), and use these to address the issues outlined above. This is
part of a major effort currently being undertaken at the PdBI with the aim of
detecting and imaging CO emission towards $\sim20$--$30$ SMGs. 
In \S \ref{section:observations} and
\ref{section:sources} we describe the observations and derive properties for
each new source observed.  In \S \ref{section:sample} we define a sample of 12
sources consisting of all SMGs detected in CO to date, from which the average
physical properties of the bright SMG population are derived. We discuss their properties
and compare with those of other galaxy populations in \S \ref{section:comparison}.
Finally, \S \ref{section:implications} discusses the impact of our observations
on models of galaxy formation and evolution. Throughout we adopt a flat
cosmology, with $\Omega_m=0.27$, $\Omega_\Lambda=0.73$ and
$H_0=71$\,km\,${\mbox{s}^{-1}}$\,Mpc$^{-1}$ (Spergel et al.\ 2003).

\section{Observations}\label{section:observations}
In total, 11 SMGs were targeted for CO observations with PdBI 
(see Table \ref{table:observations}), drawn from five independent
submm surveys: the SCUBA Lens Survey (SMM\,J02396$-$0134 --- Smail, Ivison \&
Blain 1997); the Hubble Deep Field (SMM\,J12360$+$6210 --- Chapman et al.\ 2003a); 
the Hawaii Survey Fields, SSA\,13 and SSA\,22 (SMM\,J13120$+$4242,
SMM\,J13123$+$4239 --- Chapman et al.\ 2005; SMM\,J22174$+$0015 --- Barger, Cowie
\& Sanders 1999); The SCUBA UK 8\,mJy Survey of the Lockman Hole East and ELAIS\,N2
(SMM\,J10523$+$5722, SMM\,J10524$+$5724,
SMM\,J16363$+$4055, SMM\,J16363$+$4056, and SMM\,J16366$+$4105 --- Scott et
al.\ 2002); and a 1200-$\mu$m MAMBO survey of the same two fields
(SMM\,J16371$+$4053\footnote{This source was denoted MM\,J16371$+$4053 in Greve
et al.\ (2004b)} --- Greve et al.\ 2004b).

Deep radio imaging was used to identify radio counterparts
to all the targeted SMGs and accurately
locate these relative to optical/near-infrared reference fields (Smail et al.\ 1999; Chapman et
al.\ 2003a; Ivison et al.\ 2002). Subsequent spectroscopy with Keck/LRIS-B (Oke
et al.\ 1995) and CFHT/OSIS-V in the case of SMM\,J02396$-$0134 (Soucail et
al.\ 1999) provided UV spectroscopic redshifts for all sources (see
Chapman et al.\ 2003b, 2005).  A subset of the sample were spectroscopically observed
in the near-infrared  to provide more reliable systemic redshifts from the 
wavelengths of redshifted H$\alpha$ or [O{\sc iii}]\,$\lambda$5007{\AA} emission lines
(Swinbank et al.\ 2004) to aid in our search for redshifted CO emission.  
Here, and for the remainder of this paper, we take the systemic redshift
to be equivalent to the CO redshift.

From our sample of 11 sources, 10 were part of our statistically-complete CO survey and were
targeted during two observing campaigns, 
during the winter of 2002--03 and the summer of 2003, in good to
excellent weather. The observations were done in D configuration in 
order to maximise the sensitivity, and used five of the six available antennae, 
giving a total of 10 baselines. We stress that the survey is on-going and that the
10 sources reported here are merely those which were observed during the 2003 season.
In addition, we included an 11th source, 
SMM\,J02396$-$0134, for which data was originally obtained by J.-P.\ Kneib and G.\ Soucail 
during summer/autumn of 1999, and these
were retrieved from the PdBI data archive. Observations of this
source were done in D and C configuration. The details of
the observations of all 11 sources are summarised in Table \ref{table:observations}.

While our main goal was to detect redshifted CO emission in the 3-mm waveband, we also used the 1.3-mm
receivers to attempt to determine accurate continuum positions and
fluxes or, in cases where a higher-$J$ CO line would coincide with the 1.3-mm
band, to search for emission from this transition, e.g.\ SMM\,J04431$+$0210 (Neri et
al.\ 2003).  To achieve our goals the correlators were configured for line and continuum
observations and simultaneously covered 580\,MHz in the 3-mm and 1.3-mm bands:
this corresponds to a typical velocity coverage of 1700 and 750\,\kms at 3 and 1.3\,mm,
respectively. 

Where a near-infrared  spectroscopic redshift was unavailable  we have
to correct for the likely systematic blueshifts of UV line features relative to
the CO emission. We therefore tuned the 3-mm receivers slightly redward of the measured
spectroscopic redshift. This meant that for a typical redshift uncertainty of
$\Delta z = 0.005$, any source at $z \ge 1$ would have its CO line peak fall
within $0.5 \times 580\,\mbox{MHz} = 290\,\mbox{MHz}$ of the 3-mm band centre.  In
a few instances, a line was detected at the edge of the band pass. The
frequency setting was then adjusted to centre the line in the bandpass, and the
source was re-observed. A source was typically observed for 2--3 tracks
(10--18\,hrs).
If no signal had been detected after this, the source was not pursued further.

All data reduction employed the IRAM {\sc gildas} software (Guilloteau \& Lucas 2000).  
This involved careful monitoring of the quality of
the data throughout a track, and subsequent flagging of any bad and high
phase-noise visibilities.  For passband calibration we typically used one or
more bright quasars. Phase and amplitude variations within each track were
calibrated out by inter-leaving reference observations of nearby fainter
quasars every 20\,m. In the best conditions, the typical rms phase noise per
baseline at 3\,mm was $\ls 10^{\circ}$, increasing to $\ls 40^{\circ}$ in the
worst cases.  Observations of the primary calibrators, 3C454.3, 3C345, 3C273,
and MWC\,349, were used to determine the flux scale.  Finally, naturally
weighted data cubes were created using {\sc gildas}.

%
%
\begin{table}
\scriptsize
\caption{Log of the PdBI observations for the 11 SMGs analysed in
this paper. The on-source integration time ($t_{int}$) is the observing time for the equivalent
six element array.}
\vspace{0.2cm}
\begin{center}
\begin{tabular}{llrc}
\hline
Source            & Observing Dates                        &   $t_{int}$ & Detection?  
    \\
& & (hr) & \\
\hline
\hline
SMM\,J02396$-$0134  & 1999 Jun 20, 24--26, 29, Aug 24, 27, Sep 7 &        7.4  & Y              \\
SMM\,J10523$+$5722  & 2003 Apr 25, May 06                    &        16.6  & N               \\
SMM\,J10524$+$5724  & 2003 Mar 27                            &        3.9  & N               \\
SMM\,J12360$+$6210  & 2003 Jun 01, 03, 18, Aug 30, Sep 01    &        9.4 & N                \\
SMM\,J13120$+$4242  & 2003 May 17, 22--23, 26                   &        9.2  & Y              \\
SMM\,J13123$+$4239  & 2003 Jun 23, 26, 27                      &        4.7 & N                 \\
SMM\,J16363$+$4055  & 2003 Mar 31, Apr 03, 21                &        11.2 & N                \\
SMM\,J16363$+$4056  & 2003 Aug 11, 22, 24                    &        9.6  & N              \\
SMM\,J16366$+$4105  & 2003 Apr 13--14, 23, May 1               &        14.9 & Y              \\
SMM\,J16371$+$4053  & 2003 Jul 25--26, 29 Aug 1,2, Sep 4       &        17.1 & Y              \\
SMM\,J22174$+$0015  & 2003 May 9--13, 29--30, Jun 4, Sep 12, 14  &        17.3 & Y              \\ 
\hline
\label{table:observations}
\end{tabular}
\end{center}
\end{table}

\section{The Sources}\label{section:sources}
No sources were detected at 1.3-mm, and the data were of so poor quality that even upper limits
were useless. For the remainder of this paper we shall therefore only discuss the 3-mm data. 
For each source the continuum level at 3-mm was estimated by extrapolating from its submm flux
and assuming a spectral index of $\beta=1.5$. In all cases we find
the contribution from the continuum emission to the CO line at 3-mm to be less than $\ls 5$\,per cent, and therefore negligible.
Of the 11 sources whose observations are analysed in
this paper, CO emission was reliably detected in five. 
Including the first three (out of three) detections 
from our survey from Neri et al.\ (2003) yields a detection rate of 57\% (8/14) to
date. 

The failure to detect CO in some of our sources is unlikely to be due to the optical
redshift being wrong, since the quality of the
spectra is generally high (Chapman et al.\ 2005). Similarly, the possibility that the identification is wrong, and
therefore the redshift is for the wrong object, appears to be remote given 
the excellent correspondence between the radio and submm (e.g.\ Ivison et al.\ 2002),
and the extremely small likelihood of finding a $z\sim 2$ galaxy at that position, which is
not related to the submm emission.
Instead we feel the explanation for the missing CO emission is due
to either too large a velocity offset 
between the CO and optical emission (putting the former outside
our selected correlator coverage, see \S \ref{section:zco-zopt}), a CO
line with a width comparable to our correlator bandwidth,
or simply that the sources are too
faint in CO to be detected in the given integration time. In the latter case,
the 6 non-detections can provide useful upper limits on their CO luminosity
and gas masses and as a result we have included them in the analysis
in this paper. 

%
%
\begin{figure*}
\begin{center}
\includegraphics[width=0.775\hsize,angle=0]{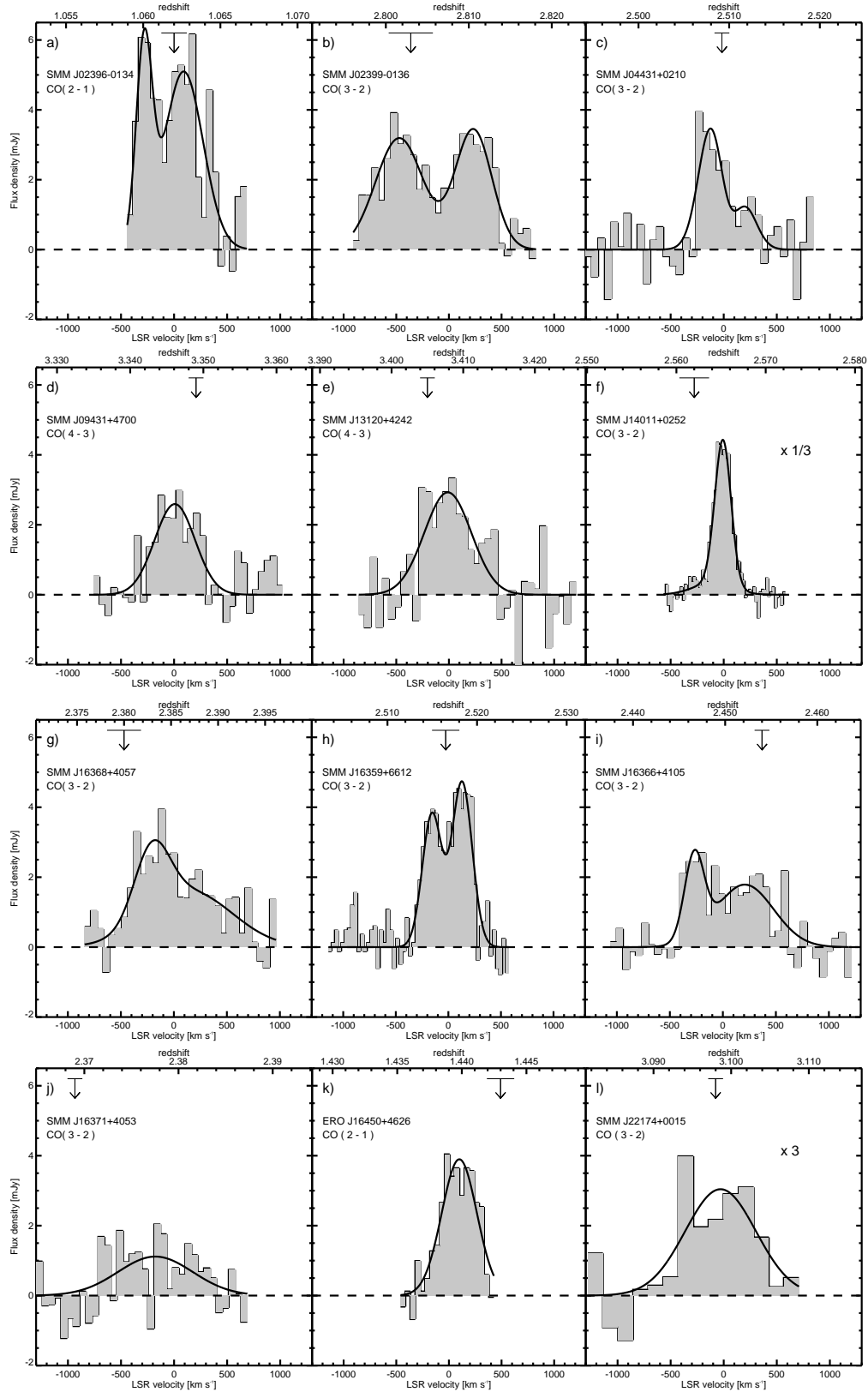}
\caption[CO line profiles of the 12 SMGs detected in CO to date.]  {CO
line profiles of the 12 SMGs detected in CO to date. The
profiles have been plotted on similar flux- and velocity-scales in order to
facilitate an easy comparison between individual sources, to achieve this the spectra
of SMM\,J14011$+$0252 and SMM\,J22174$+$0015 have been scaled by a factor
$\times 1/3$ and $\times 3$, respectively. The
LSR velocity-scale is relative to the CO redshift. Optical redshifts are
denoted by arrows and the horizontal bars on top of the arrows indicate the
uncertainty of the redshifts. The redshifts derived from the CO lines are given in
Table 2. The solid curves represent the
best-fits to the spectra using either single or double Gaussian profiles.
The spectra shown in c), d) and  g) are taken from Neri et al.\ (2003); h) is taken from
Kneib et al.\ (2004), while
b), f) and k) are from Genzel et al.\ (2003), Downes \& Solomon (2003)
and Andreani et al.\ (2000), respectively. The spectra have been binned into 
20\,MHz bins, except for f), h) and l) where the frequency binning is
7, 10 and 40\,MHz, respectively.}
\label{fig:line-profiles}
\end{center}
\end{figure*}

In remainder of this section we shall briefly describe the properties of the 
five SMGs for which new CO detections are presented in this paper as well
as the 6 non-detections.

%
%
\begin{figure*}
\begin{center}
\includegraphics[width=1.0\hsize,angle=0]{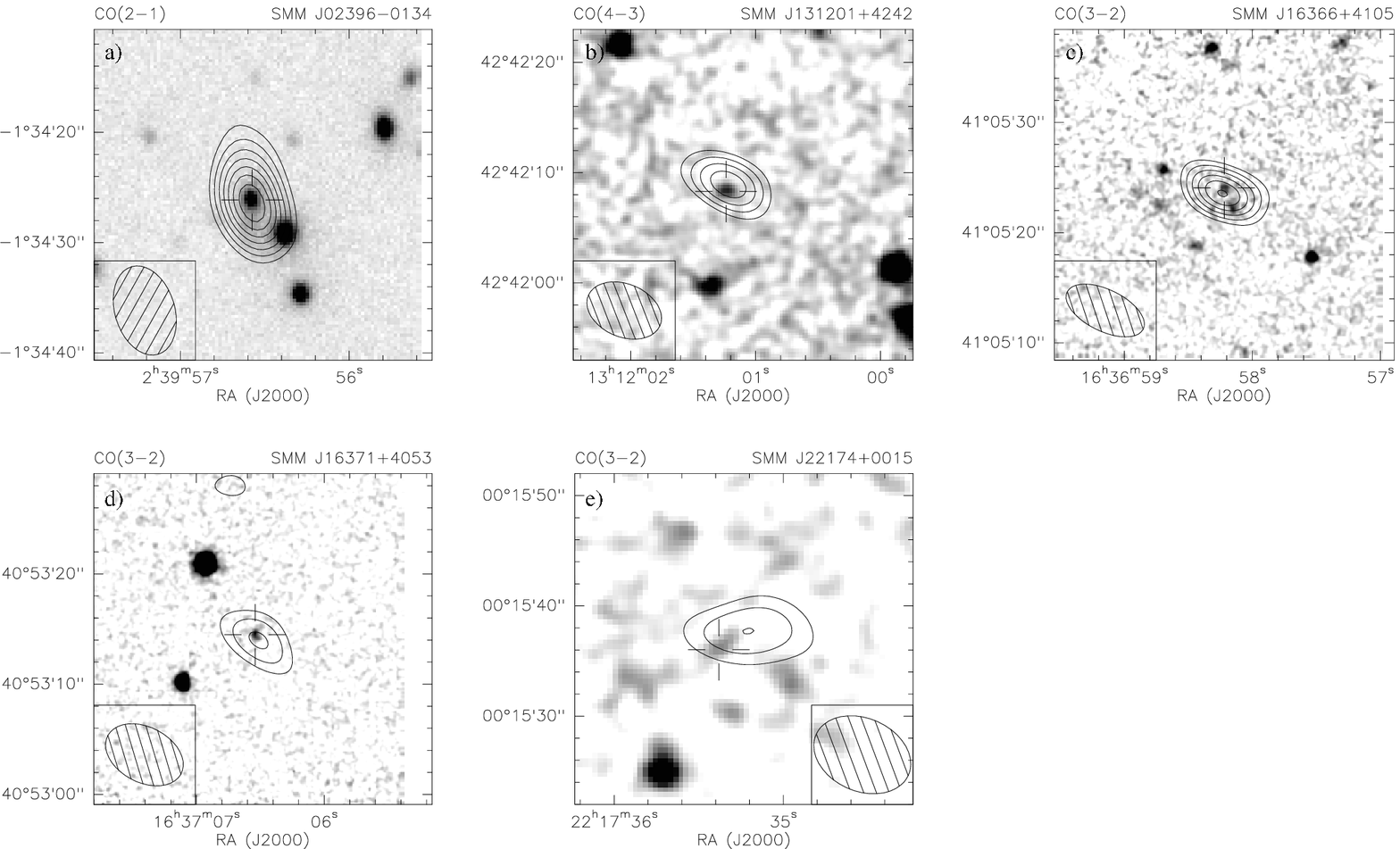}
\caption[Contours of the velocity-integrated CO emission overlaid on $K$-band
images of the five new CO-detected SMGs presented in this paper.]  
{Contours of the velocity integrated CO emission overlaid on $30\arcsecs \times 30\arcsecs$ $K$-band images of the five new
CO-detected SMGs presented in this paper. The $K$-band images are centred on the radio positions, while the open crosses mark the
position of the optical/near-infrared counterparts for which the spectroscopic redshifts were obtained.
In all panels the contours start at
$3\sigma$ and increase in steps of $1\sigma$, 
where $\sigma = 0.40, 0.32, 0.21, 0.18,$ and $0.17$\,mJy\,beam$^{-1}$
in panels a), b), c), d), and e), respectively. The synthesized beams are (panels a) to e), shown as hatched ellipses)
$8\parcsec3\times 5\parcsec3$ at position angle $15\degs$ (east of north), $6\parcsec9\times4\parcsec8$ at $66\degs$,
$7\parcsec4\times 3\parcsec9$ at $62\degs$, $7\parcsec4\times 5\parcsec0$ at $61\degs$, and $9\parcsec0\times 6\parcsec6$
at $65\degs$. The ($1\sigma$) uncertainties on measured positions in the optical and mm frames are $\ls 0\parcsec1$ and $\ls 0\parcsec3$, respectively. The $K$-band
data for a), b) and e) were published in Smail et al.\ (2002), Smail et al.\ (2004) and Chapman et al.\ (2003a), respectively, while
c) and d) were published in Ivison et al.\ (2002).}
\label{fig:maps}
\end{center}
\end{figure*}

%
%
\begin{figure*}
\vbox to220mm{\vfil \includegraphics[width=1.0\hsize,angle=0]{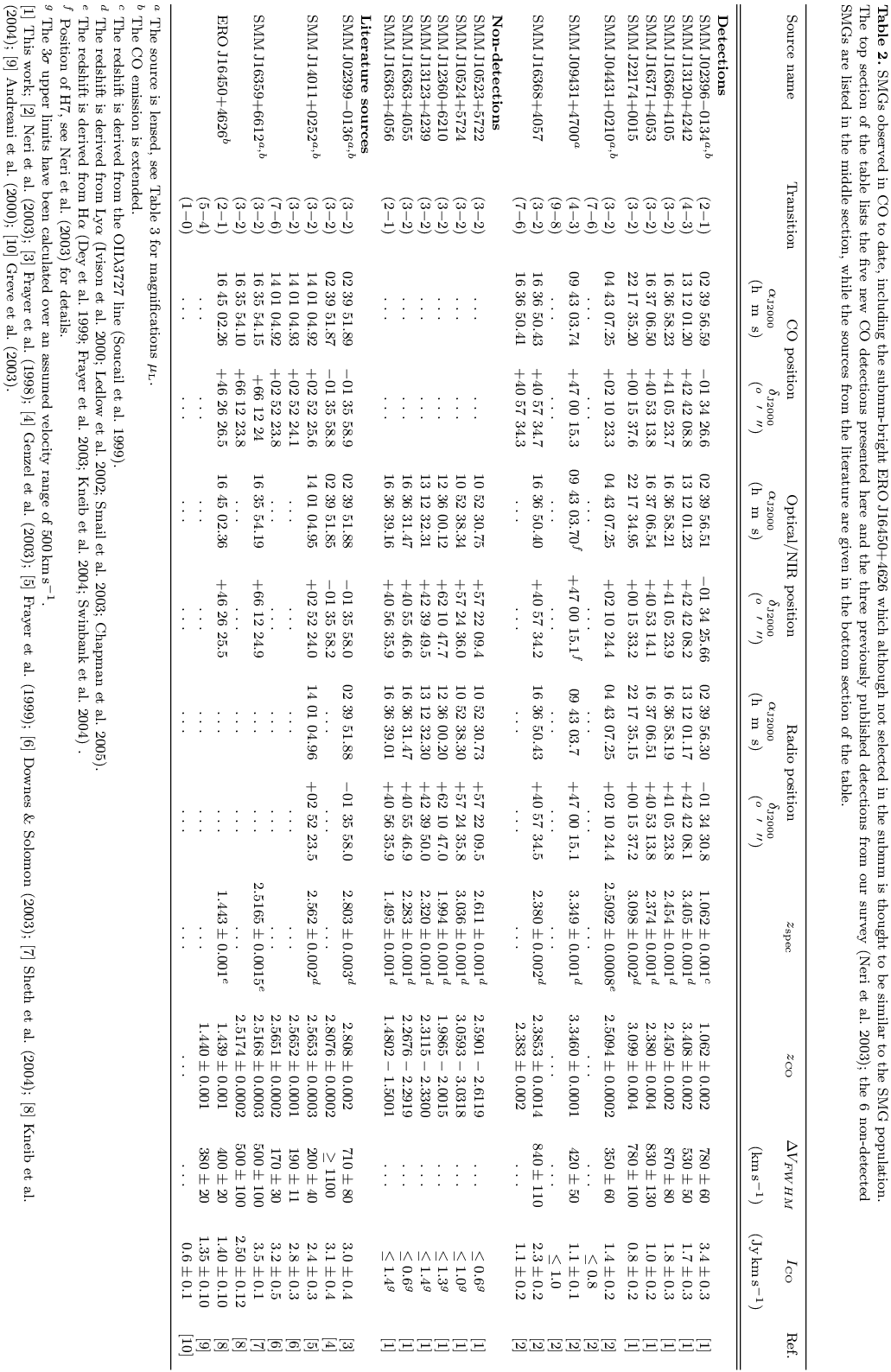}
\vfil}
\end{figure*}

\textbf{SMM\,J02396$-$0134} The \COJ{2}{1} spectrum of SMM\,J02396$-$0134 is
shown in Fig.\ \ref{fig:line-profiles}a.  The blue wing of the line is
missing due to the limited correlator bandwidth available in 1999,
but nonetheless it is evident that the line is double peaked. 
The CO redshift, $z_{\mbox{\tiny{CO}}} = 1.062\pm 0.002$, was defined as the 
flux-weighted redshift, see section \ref{section:zco-zopt} for details.
The widths of the two peaks were
estimated by simultaneously fitting two Gaussians to the spectrum, which
yielded {\sc fwhm} values $180\pm 20$\,\kms and $430\pm 50$\,\kms for the
blue-shifted and red-shifted peaks, respectively. The velocity offsets of the two peaks
were $-$280\,\kms and 90\,\kms, respectively, measured relative to the CO
redshift. From the double Gaussian fit the
velocity-integrated \COJ{2}{1} line flux is $I_{\mbox{\tiny{CO}}} = 3.4\pm
0.3$\,Jy\,km\,s$^{-1}$. A single Gaussian fit to the line 
profile leads to a {\sc fwhm} value of $780\pm 60$\,\kms. 
Due to the truncated line we can only impose a lower limit $\simeq
870$\,\kms on the full width at zero intensity ({\sc fwzi}).

The emission integrated over the line from $-460$\,\kms to 110\,\kms is shown as contours
overlaid on a $K$-band image in Fig.\ \ref{fig:maps}a.  The emission is
detected at $\ge$10$\sigma$, where $\sigma = 0.40$\,mJy\,beam$^{-1}$, and
coincides with the optical/near-infrared  counterpart to within the 
relative astrometrical
errors.  The source appears extended in the north-south direction, and an elliptical
Gaussian fit to the CO emission in the image plane yields a source size of
$9\parcsec5 \times 5\parcsec5$, which is slightly larger than the $8\parcsec3
\times 5\parcsec 3$ synthesized beam. The two velocity components of the source
have a slight spatial offset, with the blue-shifted peak (integrated from $-$460 to
$-$120\,\kms) $1.2\arcsecs$ north of the centroid of the integrated
line emission  and the red-shifted
peak (integrated from $-$120 to 110\,\kms) offset 0.3$\arcsecs$ east and
1.6$\arcsecs$ south of the central position.  The positional uncertainties in
the maps are 0.3--0.4$\arcsecs$, suggesting that the observed offsets are
significant.

\textbf{SMM\,J13120$+$4242} The \COJ{4}{3} spectrum of SMM\,J13120$+$4242 is
shown in Fig.\ \ref{fig:line-profiles}e.  A single Gaussian provides an excellent
fit to the data, as measured by the $\chi^2$ statistic, and yields 
a velocity-integrated flux density, $1.7\pm 0.3$\,Jy\,\kms and a line width of 
{\sc fwhm}\,$\simeq 530\pm 50$\,\kms. We find a systemic CO redshift 
of $z_{\mbox{\tiny{CO}}}=3.408\pm 0.002$.

The velocity-integrated \COJ{4}{3} emission yields a $\gs 6$-$\sigma$ detection
($\sigma = 0.32$\,mJy\,beam$^{-1}$) at position which is offset 0.8\arcsec\ to the north-east of the optical
counterpart (Fig.\ \ref{fig:maps}b). The positional error is $\sim1 \arcsecs$,
and so the offset is not significant. The CO emission is unresolved.

\textbf{SMM\,J16366$+$4105} The \COJ{3}{2} spectrum of SMM\,J16366$+$4105 
(Fig.\ \ref{fig:line-profiles}i) is very broad with the blue
edge of the line ending abruptly.  The spectrum exhibits a double-peaked
line similar to  SMM\,02396$-$0134, although
less distinct.  The CO redshift
is $z=2.450\pm 0.002$. A simultaneous
fit of two Gaussians provides a superior fit to the spectrum than a single Gaussian,
and places the blue-shifted and red-shifted peaks at  velocity offsets of $-270$\,km\,s$^{-1}$ and $+210$\,km\,s$^{-1}$,
respectively. The velocity-integrated line flux is $I_{\mbox{\tiny{CO}}} = 1.8\pm 0.3$\,Jy\,\kms.
A single Gaussian fitted to the spectrum yields a line width {\sc fwhm}=$870\pm
80$\,\kms -- the second largest line width of the sample. 

The integrated CO emission, shown in Fig.\ \ref{fig:maps}c, is detected at the
$\gs 8$-$\sigma$ level ($\sigma = 0.21$\,mJy\,beam$^{-1}$), and coincides with the radio and faint optical/near-infrared
counterpart. We imaged the blue-shifted and red-shifted peaks
separately, using different velocity cut-offs, in order to look for positional
offsets between the two, but found no evidence for a velocity gradient across
the source.
It is possible that part of the CO emission is related to the faint near-infrared  source $\sim
2\arcsecs$ south-west of the CO centroid, the compact radio counterpart displays
a similar faint tail of emission in this direction. SMM\,J16366$+$4105 is a prime
target for further, higher-resolution CO observations.

\textbf{SMM\,J16371$+$4053} The detection of \COJ{3}{2} towards
SMM\,J16371$+$4053 is the first for a MAMBO-selected source.  The \COJ{3}{2}
line profile is shown in Fig.\ \ref{fig:line-profiles}j. 
The flux-weighted, systemic CO redshift is 
$z_{\mbox{\tiny{CO}}} = 2.380\pm 0.004$. While clearly
detected, the poor signal-to-noise ratio prevents us from defining the line
profile, except to say that the line appears broad with {\sc fwzi}\,$\gs
900$\,km\,s$^{-1}$. A Gaussian fit yields a formal line width of {\sc
fwhm}$\sim 830$\,km\,s$^{-1}$ 
and a velocity-integrated line flux $I_{\mbox{\tiny{CO}}} = 1.0\pm 0.2$\,Jy\,\kms.  \newline
\indent The integrated emission is shown in Fig.\ \ref{fig:line-profiles}d.  The source
is detected at $\gs 5\sigma$ ($\sigma = 0.18$\,mJy\,beam$^{-1}$) and coincides with the optical/near-infrared
counterpart to within the positional errors.

\textbf{SMM\,J22174$+$0015} The \COJ{3}{2} spectrum of this source shown in
Fig.\ \ref{fig:line-profiles}l is the weakest of our detections. After heavily
binning the spectrum, we estimate the line to be significant at the
$\gs4$-$\sigma$ level.  We measure the flux-weighted, systemic CO redshift as 
$z_{\mbox{\tiny{CO}}} =3.099\pm 0.004$.
A Gaussian fit to the spectrum
yields a line width of {\sc fwhm}\,$=780\pm 100$\,\kms
and a line flux density of $I_{\mbox{\tiny{CO}}} = 0.8 \pm 0.2$\,Jy\,\kms.\newline
\indent Contours of the velocity-integrated CO emission are shown on a $K$-band image
in Fig.\ \ref{fig:maps}e. The CO emission is detected at the $\sim4.7$-$\sigma$
level ($\sigma = 0.17$\,mJy\,beam$^{-1}$), at a position offset by 
$\sim 2.5\arcsecs$ north-west from the optical counterpart, but only
$\sim 0.8 \arcsecs$ away from the radio position (phase centre).

\textbf{Non-detections} In Table 2 we list the $3\sigma$ upper limits 
on the 3-mm CO line flux spatially coincident with the radio position 
as well as the redshift range searched for each of the 6 non-detections.
The upper limits were calculated using the equation $I_{\mbox{\tiny{CO}}} =
3 \sigma \left ( \delta v \Delta v_{\mbox{\tiny{FWHM}}} \right ) ^{1/2}$, where $\sigma$
is the channel-to-channel rms noise, $\delta v$ the velocity resolution and $\Delta v_{\mbox{\tiny{FWHM}}}$
the line width (see e.g.\ Seaquist, Ivison \& Hall 1995). In our calculation of the upper
limits we used the rms noise in the spectra binned to $100$\,MHz resolution and 
adopted a line width of $500$\,km\,s$^{-1}$. The integrated CO (3$-$2) map of SMM\,J12360$+$6210 shows a
$\sim 4\sigma$ peak slightly eastwards of the phase-centre, and is thus possibly another tentative detection,
albeit marginal. In the remaining cases no line emission is observed at a significance $\ge 3\sigma$.

\section{Sample Properties}\label{section:sample}

\subsection{The sample}

The main goal of this paper is to assemble a large sample of SMGs observed
in CO. In order to achieve this, 
we add to the 11 new sources in Table \ref{table:observations}
from our survey (detections as well as non-detections) and
the three earlier CO detections 
from this survey described in Neri et al.\ (2003), the
three SMGs detected in CO prior to our survey: SMM\,J02399$-$0136 (Frayer et
al.\ 1998; Genzel et al.\ 2003), SMM\,J14011$+$0252 (Frayer et al.\ 1999;
Downes \& Solomon 2003) and SMM\,J16359$+$6612 (Sheth et al.\ 2004; Kneib et al.\ 2004).  
The latter is a triply-imaged lensed galaxy with CO detected towards all
three components, and in Fig.\ \ref{fig:line-profiles}h we have displayed the combined 
(and therefore highest signal-to-noise) spectrum of all three components as given in
Kneib et al.\ (2004). For the velocity-integrated line flux we have adopted the value of the
component with the strongest detection, i.e.\ component B (Kneib et al.\ 2004) -- see Table 2.   

We also include the extremely red object, HR\,10, at
$z= 1.443$ (ERO\,J16450$+$4626; Hu \& Ridgway 1994). Initially studied
because of its extremely red optical--near-infrared  colours, 
subsequent observations with SCUBA and
{\it Hubble Space Telescope} revealed that HR\,10 is a powerful SMG with a distorted 
rest-frame UV morphology (e.g.\ Cimatti et al.\ 1998; Dey et al.\ 1999).  Hence,
HR\,10 bears all the characteristics of a SMG and would have been detected in any
of the SCUBA surveys to date. The \COJ{5}{4}, \COJ{2}{1} and \COJ{1}{0} lines
have all been detected towards this source (Andreani et al.\ 2000; Greve et
al.\ 2003).

Finally, we note that Hainline et al.\ (2004) detected \COJ{3}{2} towards
the submm source SMM\,J04135$+$10277, the first type-1 QSO to be
selected at submm wavelengths (Knudsen, van der Werf \& Jaffe 2003), as part of an Owens
Valley Millimeter Array survey of CO emission towards high-redshift
QSOs. SMM\,J04135$+$10277 is not only one of the brightest submm sources known
but also an extremely luminous CO source. While SMM\,J04135$+$10277 was
discovered in a blank-field (albeit lensing-assisted) submm survey and thus
meets the selection criteria for our sample, only about 3 per cent of
radio-identified SMGs are broad-line QSOs (Chapman et al.\ 2005), suggesting
that this source is not representative of the SMG population as a whole.
SMM\,J04135$+$10277 is therefore not included in our sample of CO-detected
SMGs.

The final SMG sample consists of 12 CO detections and 6 non-detections, their 
observational properties are listed in Table 2. The CO line profiles of 
the 12 detected sources are shown in Fig.\ \ref{fig:line-profiles}.
In the following sections we shall use the CO observations to investigate the
physical properties of the SMG population, see Table \ref{table:sample-average}.
In the case where a SMG is known to be gravitationally-lensed, we have
corrected the luminosities, gas masses, and linear sizes, etc., using the
gravitational lensing magnification factors listed in Table \ref{table:sample-average}.

\subsection{Comparison of optical and CO redshifts}\label{section:zco-zopt}

In general, CO lines provide excellent systemic redshifts for galaxies, by
tracing the extended molecular gas distribution rather than the ionised gas
produced by shocked outflows or accretion onto an AGN. 
The latter is typically
traced by high-ionisation, e.g.\ broad C\,{\sc iv}$\lambda$1549 lines, which have been shown to be
systematically blue-shifted with respect to the systemic redshift in quasars
(e.g.\ Richards et al.\ 2002), and by the Ly$\alpha$ line, which 
in the majority of high-redshift CO detections is blue-shifted
with respect to the CO redshift due to outflows and dust obscuration
(e.g.\ Hainline et al.\ 2004).
The purpose of this section is to determine if similar
systematic offsets between the CO and Ly$\alpha$ redshifts 
exist within the SMG population. We have measurements of
the Ly$\alpha$ redshifts for
eight of the CO-detected SMGs, see Table 2.

A large fraction of CO line profiles shown in 
Fig.\ \ref{fig:line-profiles} are double peaked or asymmetric, making it difficult to
accurately determine the CO redshift by fitting Gaussian profiles to the
spectra. Instead, the CO redshift of a source was determined by computing its flux-weighted
redshift, i.e.\ $z_{\mbox{\tiny{CO}}} = \sum I(z)z/\sum I(z)$. The
error on the redshift is given by $\Delta z^2_{\mbox{\tiny{CO}}} = \sum I(z) (z - z_{\mbox{\tiny{CO}}})^2
/\sum I(z)$. The CO redshifts computed in this way are listed in Table 2.

\begin{figure}
\begin{center}
\includegraphics[width=1.0\hsize,angle=0]{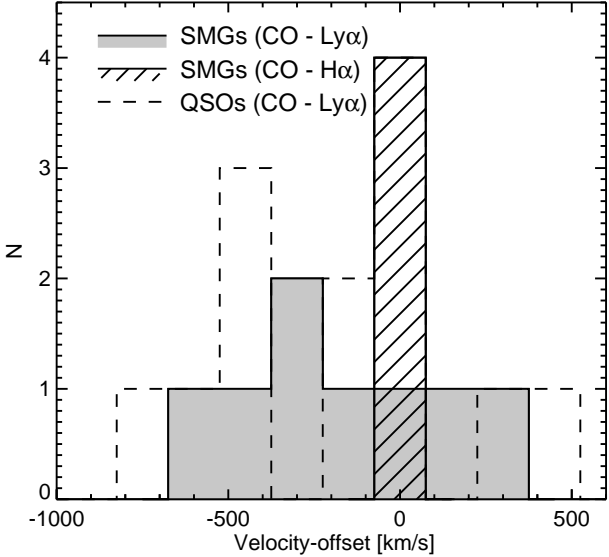}
\caption[The distribution of velocity offsets obtained
from CO line observations and restframe-UV spectroscopy using the Ly$\alpha$ line.]
{The histograms show the distribution of velocity differences derived 
from the redshifts from CO line observations and restframe-UV spectroscopy. We
show the distribution for the
eight SMGs for which  spectroscopic redshifts have been derived
from the Ly$\alpha$ line (grey shaded histogram) and for
the 15 QSOs in Table \ref{table:all-co-detections} (two of the QSOs have
velocity-offsets $\ls -1000$\,km\,s$^{-1}$). 
The hashed histogram compares
the redshifts between the CO and H$\alpha$ emission line for
the 4 SMGs with redshifts derived from H$\alpha$ (3) or [O{\sc ii}]\,$\lambda$3727\,{\AA} (1).
The bin size is 150\kms which also reflects the typical error in the
velocity offsets.}
\label{figure:zCO-zopt}
\end{center}
\end{figure}

In Fig.\ \ref{figure:zCO-zopt} we have plotted the distribution of 
velocity offsets corresponding to the differences between the CO and Ly$\alpha$ redshifts
for the SMG sample.
Negative velocity-offsets correspond to blue-shifted UV emission with respect to the 
systemic CO redshift. Also shown are the velocity offsets for the
four SMGs with H$\alpha$ redshifts.

%
%

The velocity offsets between the Ly$\alpha$ and CO redshifts
of SMGs show a tendency for blue-shifted
offsets (five and two SMGs have blue- and red-shifted offsets, respectively, with one source
consistent with no velocity offset within the errors, $\sim 150$\kms), although
this needs to be confirmed once a larger sample of SMGs is studied. 
The 15 QSOs detected in CO to date show
a similar tentative excess of negative CO-Ly$\alpha$ velocity offsets (Fig.\ \ref{figure:zCO-zopt}), although in this
case the the shift is due to strong absorption of the line -- most likely from gas
on small scales near the nucleus. In the case of SMGs, the offsets are
likely to come from absorption from a wind on larger kpc-scales -- where we only
see the line scattered from the far side of the wind/shell due to the absorption
of the blueshifted component.

It is clear from Fig.\ \ref{figure:zCO-zopt} that in the majority of the SMGs,
significant velocity offsets exist between the Ly$\alpha$ and CO emission. 
Such large offsets illustrate how easy
it would be to miss a CO line based on blind tuning to the restframe-UV redshift,
and highlight the need for future submm/mm correlators with large instantaneous
bandwidths.

Realising the potential danger of not detecting some of the SMGs because of
large offsets between the CO and restframe-UV redshifts, we have undertaken 
a programme to target
SMGs using the near-infrared  spectrographs NIRSPEC on Keck-II, OHS on Subaru and
ISAAC on the VLT, to allow us to measure redshifts from 
restframe optical emission lines, 
which should provide more reliable estimates of the systemic
redshifts (e.g.\ Simpson et al.\ 2004; Swinbank et al.\ 2004).
The usefulness of this approach is demonstrated by the hashed histogram
in Fig.\ \ref{figure:zCO-zopt} which shows that for the four SMGs with
redshifts derived from H$\alpha$ (or [O{\sc ii}]\,$\lambda$3727\,{\AA} in the case of SMM\,J02396$-$0134 -- Soucail et al.\ 1999), 
the velocity offsets amount to no more than $\pm 75$\kms.

\subsection{CO Luminosities and gas masses}\label{section:CO-H2}

The CO line luminosities, $L'_{\mbox{\tiny{CO}}}$, of the individual SMGs 
were derived from their velocity-integrated line flux densities following
Solomon et al.\ (1997), and corrected for gravitational lensing if
necessary (see Table \ref{table:sample-average}).  For the eight sources detected in \COJ{3}{2} we
find a median luminosity of $\langle \LOJ{3}{2} \rangle = (3.8 \pm 2.3)\times
10^{10}$\,\LCOunits.  Two sources were detected in \COJ{4}{3}
(SMM\,J09431$+$4700 and SMM\,J13120$+$4242) and two in \COJ{2}{1}
(SMM\,J02399$-$0134 and ERO\,J16450$+$4626). Note
that for HR\,10/ERO\,J16450$+$4626 we have used the \COJ{2}{1} spectrum
(Andreani et al.\ 2000) rather than the \COJ{1}{0} measurement by Greve et
al.\ (2003) which lacks velocity information.

Assuming intrinsic velocity/area-averaged brightness temperature line ratios
corresponding to an optically thick, thermalised gas, i.e.\ $r_{32} =
 T_b$(3$-$2)/$T_b$(1$-$0) = $L'$(3$-$2)/$L'$(1$-$0) = 1, and similarly $r_{43}
= 1$ and $r_{21} = 1$, we derive a median \COJ{1}{0} luminosity 
of $\langle \LOJ{1}{0} \rangle = (3.8 \pm 2.0)\times 10^{10}$\,\LCOunits, where
we have included all sources, except for the non-detections. If we assume more realistic values of
$r_{32} = 0.64$ (Devereux et al.\ 1994), $r_{43} = 0.45$ (Papadopoulos et
al.\ 2000), and $r_{21} = 0.9$ (e.g.\ Braine \& Combes 1992; Aalto et al.\ 1995)
as derived from ISM studies in local starburst galaxies, we instead find a median
\COJ{1}{0} luminosity of $\langle \LOJ{1}{0}\rangle = (5.9 \pm 3.6)\times 10^{10}$\,\LCOunits.

\begin{figure}
\begin{center}
\includegraphics[width=1.0\hsize,angle=0]{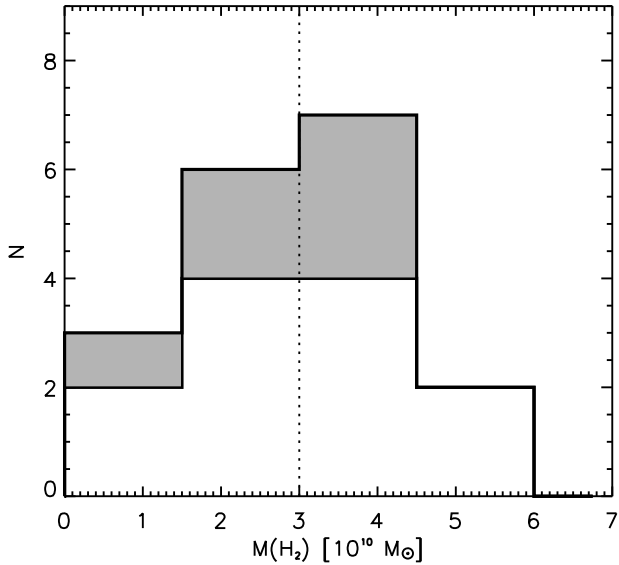}
\caption[The distribution of molecular gas masses of SMGs.]
{The distributions of molecular gas masses of the SMG sample assuming a
conversion factor of \XCO=0.8\,(K\,km\,s$^{-1}$\,pc$^2$)$^{-1} \Msolar$ and
line ratios $r_{32}=r_{43}=1$. Non-detections are
included as the shaded part of the histogram.
The median gas mass of the sample, not including the non-detections, 
is $\langle M(\mbox{H}_2)\rangle = (3.0\pm 1.6) \times 10^{10}\,\Msolar$ 
(dotted vertical line). }
\label{fig:gas-masses}
\end{center}
\end{figure}

The observed CO luminosities are converted into molecular gas masses using
$M(\mbox{H}_2) = \XCO \LOJ{1}{0}$, where \XCO\,$ =
0.8$\,(K\,km\,s$^{-1}$\,pc$^2$)$^{-1} \Msolar$ is the conversion factor
appropriate for UV-intense environments, as derived from observations of local
ULIRGs (Downes \& Solomon 1998). The molecular gas masses derived for each
source are given in Table \ref{table:sample-average} for assumed (2$-$1)/(1$-$0),
(3$-$2)/(1$-$0) and (4$-$3)/(1$-$0) line ratios of unity.  The distribution of SMG
molecular gas masses is shown in Fig.\ \ref{fig:gas-masses}, from which we
derive a median value $\langle M(\mbox{H}_2)\rangle = (3.0\pm 1.6) \times 10^{10}\,\Msolar$
(not including the upper limits from the non-detections).
Given the modest size of our sample we are not in a position to separate the observed
scatter in the distribution of molecular gas masses into components due to
varying inclination angles, differences in excitation conditions, and
underlying scatter in the mass distribution of SMGs.

Since these molecular gas mass estimates are based on observations of high-$J$
CO transitions with $J \geq 2$ and made assuming a thermalised, optically thick
gas, it is possible that significant amounts of cold, possibly sub-thermal,
molecular gas, could be present, but only detectable in lower CO transitions
(e.g.\ Papadopolous et al.\ 2001; Papadopoulous \& Ivison 2002; Greve et
al.\ 2003). Furthermore, if metal-poor gas is present, the gas mass could be higher.
Finally, CO is primarily a tracer of material in the diffuse ISM and it is
possible that a significant amount of dense and clumpy gas ($\gs
10^4\,\mbox{cm}^{-3}$) is missed by our observations (e.g.\ Carilli et al.\ 2004).  The quoted H$_2$ masses in
Table \ref{table:sample-average} should therefore be considered as lower limits
to the total amount of molecular gas. 

Our estimates of the molecular gas mass include a non-negligible fraction of Helium 
which is accounted for in the adopted value of the conversion factor (Downes \& Solomon 1998).
The amount of neutral gas (H{\sc i}), however, is particular difficult to determine and significant uncertainty is associated
with $M(\mbox{H{\sc i}})$-estimates at high redshift (e.g.\ De Breuck et al.\ 2003a).
Estimates of the $\mbox{H}_2$-to-$\mbox{H{\sc i}}$ mass ratio in infrared-luminous {\em IRAS} galaxies
range from $\sim0.5$ in systems with $L_{\mbox{\tiny{FIR}}}\sim 10^{10-11}\,\Lsolar$ 
(Andreani et al.\ 1995) to $\sim 4$ for $L_{\mbox{\tiny{FIR}}}\sim 10^{12}\,\Lsolar$ 
systems (Mirabel \& Sanders 1989). The latter value may apply
to the extremely far-infrared luminous SMGs. However, 
most of the neutral gas in local galaxies is on very large scales ($\gs 10$\,kpc) 
and thus may not be relevant for our estimates.

Finally, we have looked for evidence of a correlation between molecular gas
content and redshift. Such a trend would be 
indicative of evolution within the redshift range spanned by the SMG population. 
If we only consider the 13 sources
observed in CO as part of our statistically-complete survey (i.e.\ not including
SMM\,J02396$-$0134) and split the sample into low- and high-redshift halves at the median
redshift of the sample, $<\! z\! > = 2.4$, we find that the CO detection rate in the
high-redshift ($z\ge 2.4$) subset is 5/7 (71 per cent) compared to just
2/6 (33 per cent) for sources at $z< 2.4$. The higher detection rate at 
$z\ge 2.4$ could be indicative of evolution over the redshift range $z\sim1$--$3$
in the gas masses associated with the most luminous starbursts. Using
a Cox/Hazard survival analysis, we find that the formal likelihood of 
a variation in detection rate with redshift in the 13 sources observed by our survey is
$P=0.995$, suggestive of a real trend. If we instead include all the sources
listed in Table 2 the trend weakens (the detection rates become
4/8 and 8/10 for the low- and high-redshift bins, respectively). 
This is unsurprising since the bias against
publishing non-detections in the literature means that this comparison
is less reliable than that for our statistically-complete sample.

%
%
\setcounter{table}{2}
\begin{table*}
\scriptsize
\caption{Physical properties derived from the CO observations.}
\vspace{0.2cm}
\begin{center}
\begin{tabular}{lccccccc}
\hline
Source & $\mu_{\mbox{\tiny{L}}}^{a}$ & $D(1\arcsecs)^{b}$ & Transition & $L'_{\mbox{\tiny{CO}}}$$^{b}$          & $M(\mbox{H}_2)$$^{b,c}$    & $\Delta V_{\mbox{\tiny{FWHM}}}^d$   &  $M_{\rm dyn}^{e}$\\
       &         & (kpc)         &   & ($\times 10^{10}$\,K\,km\,s$^{-1}$pc$^2$) & ($\times 10^{10}$\,$\Msolar$) & (km\,s$^{\tiny -1}$) & $(\times 10^{11} \Msolar)$\\
\hline
\hline
SMM\,J02396$-$0134 & 2.5 & 3.3  & (2$-$1)  & $2.1\pm 0.2$    & $1.7\pm 0.2$ & $780\pm 60$   & 1.2 \\
SMM\,J02399$-$0136 & 2.5 & 3.2  & (3$-$2)  & $4.8\pm 0.8$    & $3.8\pm 0.6$ & $1360\pm 50$  & 3.5  \\
SMM\,J04431$+$0210 & 4.4 & 1.9  & (3$-$2)  & $1.1\pm 0.2$    & $0.9\pm 0.2$ & $350\pm 60$   & 0.1  \\
SMM\,J09431$+$4700 & 1.2 & 6.3  & (4$-$3)  & $2.7\pm 0.3$    & $2.2\pm 0.2$ & $420\pm 50$   & 0.7  \\
SMM\,J10523$+$5722 & 1.0 &  8.1 & (3$-$2)  & $\le 2.1$    & $\le 1.4$ & ...   &  ...\\
SMM\,J10524$+$5724 & 1.0 &  7.8  & (3$-$2)  & $\le 4.6$    & $\le 3.0$ & ...   &  ... \\
SMM\,J12360$+$6210 & 1.0 &  8.5 & (3$-$2)  & $\le 2.9$    & $\le 1.6$ & ...   &  ... \\
SMM\,J13120$+$4242 & 1.0 & 7.5  & (4$-$3)  & $5.3\pm 0.9$    & $4.2\pm 0.7$ & $530 \pm 50$  & 1.2  \\
SMM\,J13123$+$4239 & 1.0 & 8.3  & (3$-$2)  & $\le 4.1$    & $\le 2.4$ & ...   &  ... \\
SMM\,J14011$+$0252$^f$ & 5.0 & 1.6  & (3$-$2)  & $1.9\pm 0.2$    & $1.5\pm 0.2$ & $190\pm 11$   & 0.03  \\
SMM\,J16368$+$4057 & 1.0 & 8.3  & (3$-$2)  & $7.0\pm 0.6$    & $5.6\pm 0.5$ & $840\pm 110$  & 3.5  \\
SMM\,J16359$+$6612 &  22 & 0.4   & (3$-$2)  & $0.4\pm 0.2$    & $0.3\pm 0.2$ & $500\pm 100$ & 0.06 \\
SMM\,J16363$+$4055 & 1.0 &  8.3 & (3$-$2)  & $\le 1.7$    & $\le 0.9$ & ...   & ...  \\
SMM\,J16363$+$4056 & 1.0 &   8.5 & (2$-$1)  & $\le 4.1$    & $\le 1.8$ & ...   & ...  \\
SMM\,J16366$+$4105 & 1.0 & 8.2  & (3$-$2)  & $5.7\pm 1.0$    & $4.6\pm 0.7$ & $870\pm 80$   & 3.7  \\
SMM\,J16371$+$4053 & 1.0 & 8.3  & (3$-$2)  & $3.0\pm 0.9$    & $2.4\pm 0.7$ & $830\pm 130$  & 3.4  \\
ERO\,J16450$+$4626 & 1.0 & 8.5  & (2$-$1)  & $3.8\pm 0.3$    & $3.0\pm 0.2$ & $400\pm 20$   & 0.8  \\
SMM\,J22174$+$0015 & 1.0 & 7.8  & (3$-$2)  & $3.8\pm 1.0$    & $3.0\pm 0.7$ & $780\pm 100$  & 2.8  \\
\noalign{\smallskip}
Median$^g$         &  & $7.5\pm 3.1$  &  & $3.8\pm 2.0$    & $3.0\pm 1.6$ & $780\pm 320$  & $1.2\pm 1.5$  \\
\hline
\end{tabular}
\end{center}
\vspace*{-0.2cm} 
\hspace*{-7.7cm}$^a$ Assuming equal flux magnification and linear magnification.\\
\hspace*{-2.18cm}$^b$ $D(1\arcsecs)$ is the linear distance in kpc corresponding to $1\arcsecs$ at the given redshift and corrected for the lensing amplification $\mu_{\mbox{\tiny{L}}}$. \\
\hspace*{-8.10cm}$^c$ Derived assuming \XCO=0.8\,(K\,km\,s$^{-1}$\,pc$^2$)$^{-1} \Msolar$. \\
\hspace*{-6.80cm}$^d$ The line widths are derived from a single Gaussian fit to the line profile.\\
\hspace*{-8.88cm}$^e$ Calculated adopting a source diameter of 0.5\arcsec. \\
\hspace*{-2.9cm}$^f$ The amplification factor for this source was recently revised from 2.5 to $\sim 5$ (see discussion in Swinbank et al.\ 2004). \\
\hspace*{-7.39cm}$^g$ The non-detections (upper limits) are not included in the median.
\label{table:sample-average}
\end{table*}

\subsection{Line widths and dynamical masses}\label{section:lwdm}

A striking feature of the SMG population is their
typically broad CO line profiles, Table \ref{table:sample-average}. 
The median observed {\sc fwhm} of the sample
is $780\pm 320$\,\kms and the median {\sc fwzi} is
$850$\,\kmseol. These values do not take into account any geometrical effects,
such as inclination angle, which could effect the 
individual estimated line widths by more than a factor of two.  For example, in the case of SMM\,J14011$+$0252 it is
plausible that the relatively small line width is due to the fact that the
angular momentum vector of the system is
closely aligned to our line-of-sight (Tecza et al.\ 2004). Hence the observed
{\sc fwhm} are firm lower-limits to the true orbital velocities and assuming our sample has
random orbital inclinations, then the average correction should be $\sec \pi/4=1.4\times$.

In order to use our observed CO line widths to derive the dynamical mass of the SMGs 
we must know the spatial extent of the CO. 
However, so far CO emission has been reliably resolved in only one SMG: SMM\,J02399$-$0136 (Genzel et
al.\ 2003). While a number of SMGs show tentative evidence of extended CO emission
(Ivison et al.\ 2001; Neri et al.\ 2003; Greve et al.\ 2003), their low signal-to-noise ratios 
prevent a robust determinations of the source size. 
Instead, we have assumed a conservative source diameter
of $\sim 0.5\arcsec$, which corresponds to a median linear diameter  of 
$3.7$\,kpc (correcting for gravitational lensing where relevant), 
see Table \ref{table:sample-average}. This is a factor of
two smaller than the source sizes estimated by Neri et al.\ (2003), but 
consistent with characteristic size estimates from recent high-resolution radio observations 
of SMGs with MERLIN by Chapman et al.\ (2004), and with results from 
high-resolution PdBI CO observations (Tacconi et al.\ 2005) of the 
three SMGs detected in CO by Neri et al.\ (2003).  

Another notable feature of our sample is the high fraction of SMGs which show evidence
of double-peaked CO line profiles. Four sources (SMM\,J02396$-$0134, SMM\,J02399$-$0136, SMM\,J16359$+$6612 and
SMM\,J16366$+$4105) show unambiguous double-peaked profiles, and a further 
two sources (SMM\,J04431$+$0210 and SMM\,J16368$+$4057) are 
better fit by a double Gaussian than a single Gaussian, as measured
by the reduced $\chi^2$ of the fit. In the case of SMM\,J16368$+$4057
further evidence of multiple CO peaks comes from H$\alpha$ IFU observations (Swinbank
et al.\ in preparation) which show emission components at velocity-offets similar to those we
infer from the double Gaussian fit.
Thus, at least 4/12 (33 per cent), and possibly
as many as 6/12 (50 per cent), of the detected sample show evidence of having more than 
one CO-emitting component in their spectra. Comparing the occurrence of double-peaked
line profiles in sources as a function of redshift within our sample, we see that
the double-peaked sources appear to lie at the lower-redshift end of our sample,
5/7 at $z\le 2.5$, compared to 1/5 for the SMGs at $z> 2.5$. 
A possible explanation for this tentative trend is discussed later.

Such multi-peaked line profiles are a tell-tale sign of orbital motion
under the influence of gravity, and can usually be attributed to either a disk or a merger.
To determine the likely structure of the gas reservoir, we 
use our median gas mass ($3\times 10^{10}\,\Msolar$) and adopt a typical diameter of $\sim 4$\,kpc
for the gas reservoir, to derive a mass surface
density of  $(\Sigma\sim 2.4\times 10^3\,\Msolar\,\mbox{pc}^{-2})$ for this structure.  
This is extremely high and implies that if the gas is present in 
a disk then this will have 
a Toomre parameter $Q < 1$\footnote{Toomre's stability criterion says that $Q
= \frac{2 v_s \Omega}{\pi G \Sigma} \ge 1$ in order for a gaseous disk to be
stable: see e.g.\ Binney \& Tremaine (1987). For a surface density $\Sigma \sim
2.4\times 10^{3}\,\Msolar\,\mbox{pc}^{-2}$, $Q$ will always be \emph{less} than
one for any realistic values of the sound speed $v_s$ and circular frequency
$\Omega$.}, indicating it will be unstable to bar formation. 
The disk would then  collapse on a
timescale comparable to the sound crossing time, which is much shorter than the
expected $\sim 10$--100\,Myr duration of the SMG phase, see \S \ref{section:discussion}.
We conclude  that the kinematics responsible for the double-peaked CO lines
are unlikely to arise from gas distributed in a stable disk, instead a more
likely scenario is that they
reflect a merger of two gas-rich components or from a disk collapsing
under gravitational instability.

%
%

With almost all of the SMGs being spatially unresolved in CO we are unable to determine
whether SMGs typically contain gaseous disks in their centres or if they
are mergers. However, given their large gas mass densities, as well as
the high fraction of double-peaked profiles we feel that SMGs are more
likely to be mergers. Additional evidence for a merger origin comes from
the detection of close companions in CO in some of our fields (Neri et al.\ 2003),
as these these are likely to be interacting or merging given the small spatial and
velocity differences. Thus, dynamical masses were calculated using the 'merger'
formula given in Neri et al.\ 2003 (see also Genzel et al.\ 2003), and reported in 
Table \ref{table:sample-average}. Neri et al.\ (2003) assumed a stable disk model which gives
dynamical masses lower by a factor of two; but since they adopted a
characteristic  source
sizes twice as large as ours, the resulting dynamical mass estimates coincide.
The median SMG dynamical mass within $R\ls 2$\.kpc is $\langle M_{dyn}\rangle \simeq
(1.2\pm 1.5) \times 10^{11}\,\Msolar$, where we have 
applied an average correction of $1.4\times$, assuming random inclinations
for the sample overall.
We estimate a median molecular gas-to-dynamical mass fraction in SMGs of
$\langle M_{\rm gas}/M_{\rm dyn}\rangle \sim 0.3$ ($\sim 0.5$ if we assume 
a disk model). 
This indicates that the contribution of the gas reservoir to the dynamics
of the central regions of typical SMGs is substantial, 
and that the gas has already concentrated into the central regions
of the galaxy by some dissipative process.
It should be noted, however, that significant uncertainties
are associated with the dynamical mass estimates, especially 
if the system is not in dynamical equilibrium, or if the
observed line is not a good tracer of the dynamical state
of the galaxy (also see the discussion in Neri et al.\ 2003).  

\section{Comparison with other populations}\label{section:comparison}

Unlike optical/near-infrared  and X-ray observations, CO observations are relatively
unaffected by either dust extinction or AGN activity, and a comparison between the CO
properties of SMGs and that of other galaxy populations could therefore
potentially provide us with a clean way of comparing the bulk properties of
SMGs with other galaxy populations at both high and low redshifts; although,  
by virtue of the submm selection technique such a comparison is of course biased
towards dust-enshrouded sources.

We list in Table \ref{table:all-co-detections} \emph{all} non-SMG CO sources detected
at $z\geq 1$.  The bulk of these are extreme AGN
such as submm-bright, high-redshift radio galaxies (HzRGs) or QSOs (e.g.\ Omont
et al.\ 1996; Papadopoulos et al.\ 2000; Cox et al.\ 2002).  

Due to their similar properties, in particular their large far-infrared
luminosities, distorted morphologies and optical/near-infrared colours (e.g.\ Ivison
et al.\ 2002; Smail et al.\ 2004), SMGs are commonly thought to be high-redshift
analogues of local ULIRGs. To test this claim we now compare the CO
properties of SMGs with ULIRGs. To this end we have used the samples of Sanders
et al.\ (1991; SA91) and Solomon et al.\ (1997; SO97) which consist of 48 and 37
local ULIRGs (and less luminous LIRGs) in the redshift range $z=0.03$--$0.27$,
respectively, and the sample of 60 (U)LIRGs from Yao et al.\ (2003; Y03),
selected from the SCUBA Local Universe Galaxy Survey (SLUGS, Dunne et
al.\ 2000).

%
%
\begin{table*}
\scriptsize
\caption{List of published high-redshift CO detections of LBGs, QSOs and HzRGs. The
sources are sorted according to their redshifts.
This table, together with Table 2summarises the 
complete list of CO detections of sources at $z\geq 1$ to date. The
velocity-integrated line fluxes, $I_{\mbox{\tiny{CO}}}$, have not
been corrected for gravitational amplification.}
\vspace{0.2cm}
\begin{center}
\begin{tabular}{llccccll}
\hline
Source & Type & Transition & $z_{\rm spec}$ & $z_{\mbox{\tiny{CO}}}$ & $\Delta V_{\rm FWHM}$ & $I_{\mbox{\tiny{CO}}}$       & Ref.  \\
       &      &            &            &                        & (km\,s$^{-1}$)    &  (Jy\,km\,s$^{-1}$)          &            \\ 
\hline
\hline
MS\,1512$-$cB\,58$^a$	 & LBG     & (3$-$2)  & $2.727$               & $2.7265\pm 0.0004$ & $174 \pm 43$    & $0.37 \pm 0.08$  & [1] \\
\hline
Q\,0957$+$561$^a$        & QSO     & (2$-$1)  & $1.413$               & $1.414$              & $440$           & $1.2$            & [2],[3]\\
IRAS\,F10214$+$4724$^a$  & QSO     & (3$-$2)  & $2.286$  & $2.2867\pm 0.0003$              & $250$           & $21$             & [4],[5]\\
                         &         & (3$-$2)  & . . .                  & $2.2855\pm 0.0003$   & $230\pm 30$     & $4.1\pm 0.9$     & [6]\\
                         &         & (3$-$2)  & . . .                  & $2.2854\pm 0.0001$   & $220\pm 30$     & $4.2\pm 0.8$     & [7]\\
                         &         & (6$-$5)  & . . .                  & $2.2857\pm 0.0003$   & $240\pm 30$     & $9.4\pm 2.0$     & [6]\\
H\,1413$+$117$^a$	 & QSO     & (3$-$2)  & $2.5582\pm 0.0003$    & $2.558$              & $326$           & $8.1$            & [8]\\
        	         &         & (3$-$2)  & . . .                  & . . .                 & $352\pm 81$     & $14.4\pm 4.4$    & [9]\\
         	         &         & (3$-$2)  & . . .                  & $2.5579$             & $362\pm 23$     & $9.9\pm 0.6$     & [10]\\
         	         &         & (3$-$2)  & . . .                  & $2.55784\pm 0.00003$ & $416\pm 6$      & $13.2\pm 0.2$    & [11]\\
         	         &         & (4$-$3)  & . . .                  & $2.5579$             & $375\pm 16$     & $21.1\pm 0.8$    & [10]\\
         	         &         & (5$-$4)  & . . .                  & $2.5579$             & $398\pm 25$     & $24.0\pm 1.4$    & [10]\\
         	         &         & (7$-$6)  & . . .                  & $2.5579$             & $376$           & $47.3\pm 2.2$    & [10]\\
VCV\,J1409$+$5628	 & QSO     & (3$-$2)  & $2.562$               & $2.585\pm 0.001$     & $370\pm 60$     & $2.4\pm 0.7$     & [12]\\
                 	 &         & (3$-$2)  & . . .                  & $2.5832\pm 0.0001$   & $311\pm 28$     & $2.3\pm 0.2$     & [13]\\
                 	 &         & (7$-$6)  & . . .                  & . . .                 & . . .            & $4.1\pm 1.0$     & [13]\\
MG\,0414$+$0534$^a$	 & QSO     & (3$-$2)  & $2.639\pm 0.002$      & $2.639$              & $580$           & $2.6$            & [14]\\
LBQS\,1230$+$1627B       & QSO     & (3$-$2)  & $2.735\pm 0.005$      & $2.741\pm 0.002$     & . . .            & $0.80\pm 0.26$   & [15]\\
RX\,J0911$+$0551$^a$	 & QSO     & (3$-$2)  & $2.800$               & $2.796\pm 0.001$     & $350\pm 60$     & $2.9\pm 1.1$     & [12]\\
SMM\,J04135$+$1027$^a$   & QSO     & (3$-$2)  & $2.837\pm 0.003$      & $2.846\pm 0.002$     & $340\pm 120$ & $5.4\pm 1.3$        & [12]     \\
MG\,0751$+$2716$^a$	 & QSO     & (4$-$3)  & $3.200\pm 0.001$      & $3.200$              & $390\pm 38$     & $5.96\pm 0.45$   & [16]\\
APM\,08279$+$5255$^{a,b}$& QSO     & (1$-$0)  & $3.87$                & $3.9$                & . . .            & $0.150\pm 0.045$ & [17]\\
         	         &         & (1$-$0)  & . . .                  & . . .                 & . . .            & $0.22\pm 0.05$   & [18]\\
        	         &         & (2$-$1)  & . . .                  & . . .                 & . . .            & $1.15\pm 0.54$   & [17]\\
         	         &         & (4$-$3)  & . . .                  & $3.9114\pm 0.0003$   & $480\pm 35$     & $3.7\pm 0.5$     & [19]\\
         	         &         & (9$-$8)  & . . .                  & $3.9109\pm 0.0002$   & . . .            & $9.1\pm 0.8$     & [19]\\
PSS\,J2322$+$1944$^a$	 & QSO     & (1$-$0)  & $4.1108\pm 0.0005$    & $4.1192\pm 0.0004$   & $200\pm 70$     & $0.19\pm 0.08$   & [20]\\
        	         &         & (2$-$1)  & . . .                  & . . .                 & . . .            & $0.92\pm 0.03$   & [20]\\
        	         &         & (4$-$3)  & . . .                  & $4.1199\pm 0.0008$   & $375\pm 41$     & $4.21\pm 0.40$   & [21]\\
        	         &         & (5$-$4)  & . . .                  & $4.1199\pm 0.0008$   & $273\pm 50$     & $3.74\pm 0.56$   & [21]\\
BRI\,1335$-$0417$^b$	 & QSO     & (2$-$1)  & $4.398\pm 0.028$      & $4.4074\pm 0.0015$   & $420\pm 60$     & $0.44\pm 0.08$     & [22]\\
                	 &         & (5$-$4)  & . . .                    & . . .                  & . . .             & $2.8\pm 0.3$     & [23]\\
BRI\,0952$-$0115$^a$ 	 & QSO     & (5$-$4)  & $4.426\pm 0.020$      & $4.4337\pm 0.0006$   & $230\pm 30$     & $0.91\pm 0.11$   & [15]\\
BRI\,1202$-$0725$^b$     & QSO     & (2$-$1)  & $4.69$                & . . .                 &   . . .            & $0.49\pm 0.09$   & [22]\\ 
                         &         & (4$-$3)  & . . .                  & . . .                 & $280\pm 30$     & $1.5\pm 0.3$     & [24]\\ 
                         &         & (5$-$4)  & . . .                  & $4.6932\pm 0.002$    & $320\pm35$      & $2.40\pm 0.30$   & [25]\\
                         &         & (5$-$4)  & . . .                  &                      & $220\pm74$      & $2.7\pm 0.41$    & [25]\\
                         &         & (7$-$6)  & . . .                  & $4.6915\pm 0.001$    & $250-300$       & $3.1\pm 0.86$    & [24]\\
SDSS\,J1148$+$5251       & QSO     & (3$-$2)  & $6.43\pm 0.05$        & $6.418\pm 0.004$     & $320$           & $0.18\pm 0.04$   & [26]\\
                         &         & (6$-$5)  & . . .                  & $6.4187\pm 0.0006$   & $279$           & $0.73\pm 0.076$  & [27]\\
                         &         & (7$-$6)  & . . .                  & $6.4192\pm 0.0009$   & $279$           & $0.64\pm 0.088$  & [27]\\
\hline
53\,W002		         & HzRG    & (3$-$2)  & $2.390\pm 0.004$      & $2.394\pm 0.001$     & $540\pm 100$    & $1.5\pm 0.2$     & [28]\\
		         &         & (3$-$2)  & . . .                  & $2.3927\pm 0.0003$   & $420\pm 40$     & $1.20\pm 0.15$   & [29]\\
B3\,J2330$+$3927$^b$     & HzRG    & (4$-$3)  & $3.087\pm 0.004$      & $3.094$              & $500$           & $1.3\pm 0.3$     & [30]\\
TN\,J0121$+$1320         & HzRG    & (4$-$3)  & $3.516$               & $3.520$              & $700$           & $1.2\pm 0.4$     & [31]\\
6C\,1909$+$722           & HzRG    & (4$-$3)  & $3.5356$              & $3.532$              & $530\pm 70$     & $1.62\pm 0.30$   & [32]\\

4C\,41.17$^b$            & HzRG    & (4$-$3)  & $3.79786\pm 0.0024$               & $3.7958\pm 0.0008$             & $1000\pm 150$           & $1.8\pm 0.2$           & [33]\\
4C\,60.07$^b$            & HzRG    & (1$-$0)  & $3.788$               & $3.791$              & $\geq 1000$     & $0.24\pm0.03$    & [34]\\
                         &         & (4$-$3)  & . . .                  & $3.791$              & . . .            & $2.50\pm 0.43$   & [32]\\
TN\,J0924$-$2201         & HzRG    & (1$-$0)  & $5.1989\pm 0.0006$    & $5.202\pm 0.001$    & $250-400$     & $0.087\pm0.017$    & [35]\\
                         &         & (5$-$4)  & . . .                  & . . .              & $200-300$            & $1.19\pm 0.27$   & [35]\\
\hline
$^a$ The source is gravitationally lensed.\\
$^b$ The CO emission is resolved. 
\label{table:all-co-detections}
\end{tabular}

\hspace*{-1.12cm}[1] Baker et al.\ (2004); [2] Planesas et al.\ (1999); [3] Krips et al.\ (2003);
                [4] Brown \& Vanden Bout (1991); [5] Solomon et al.\ (1992a); [6] Solomon et al.\ (1992b); \\
\hspace*{-1.66cm}  [7] Downes et al.\ (1995); [8] Barvainis et al.\ (1994);
                  [9] Wilner et al.\ (1995); [10] Barvainis et al.\ (1997); [11] Weiss et al.\ (2003); [12] Hainline et al.\ (2004); \\
\hspace*{-0.66cm} [13] Beelen et al.\ (2004); [14] Barvainis et al.\ (1998); [15] Guilloteau et al.\ (1999);  [16] Barvainis et al.\ (2002); 
                  [17] Papadopoulos et al.\ (2001); [18] Lewis et al.\ (2002);\\
\hspace*{-1.645cm} [19] Downes et al.\ (1999); [20] Carilli et al.\ (2002b); [21] Cox et al.\ (2002); [22] Carilli et al.\ (2002a); [23] Guilloteau et al.\ (1997); [24] Omont et al.\ (1996);\\
\hspace*{-1.549cm} [25] Ohta et al.\ (1996); [26] Walter et al.\ (2003); [27] Bertoldi et al.\ (2003); [28] Scoville et al.\ (1997); 
                   [29] Alloin et al.\ (2000); [30] De Breuck et al.\ (2003a); \\
\hspace*{-2.835cm} [31] De Breuck et al.\ (2003b); [32] Papadopoulos et al.\ (2000); [33] De Breuck et al.\ (2005); [34] Greve et al.\ (2004a); [35] Klamer et al.\ 2005. 
\end{center}
\end{table*}

\subsection{CO luminosities and gas masses}\label{section:clagm}

%
%
\begin{figure}
\begin{center}
\includegraphics[width=1.0\hsize,angle=0]{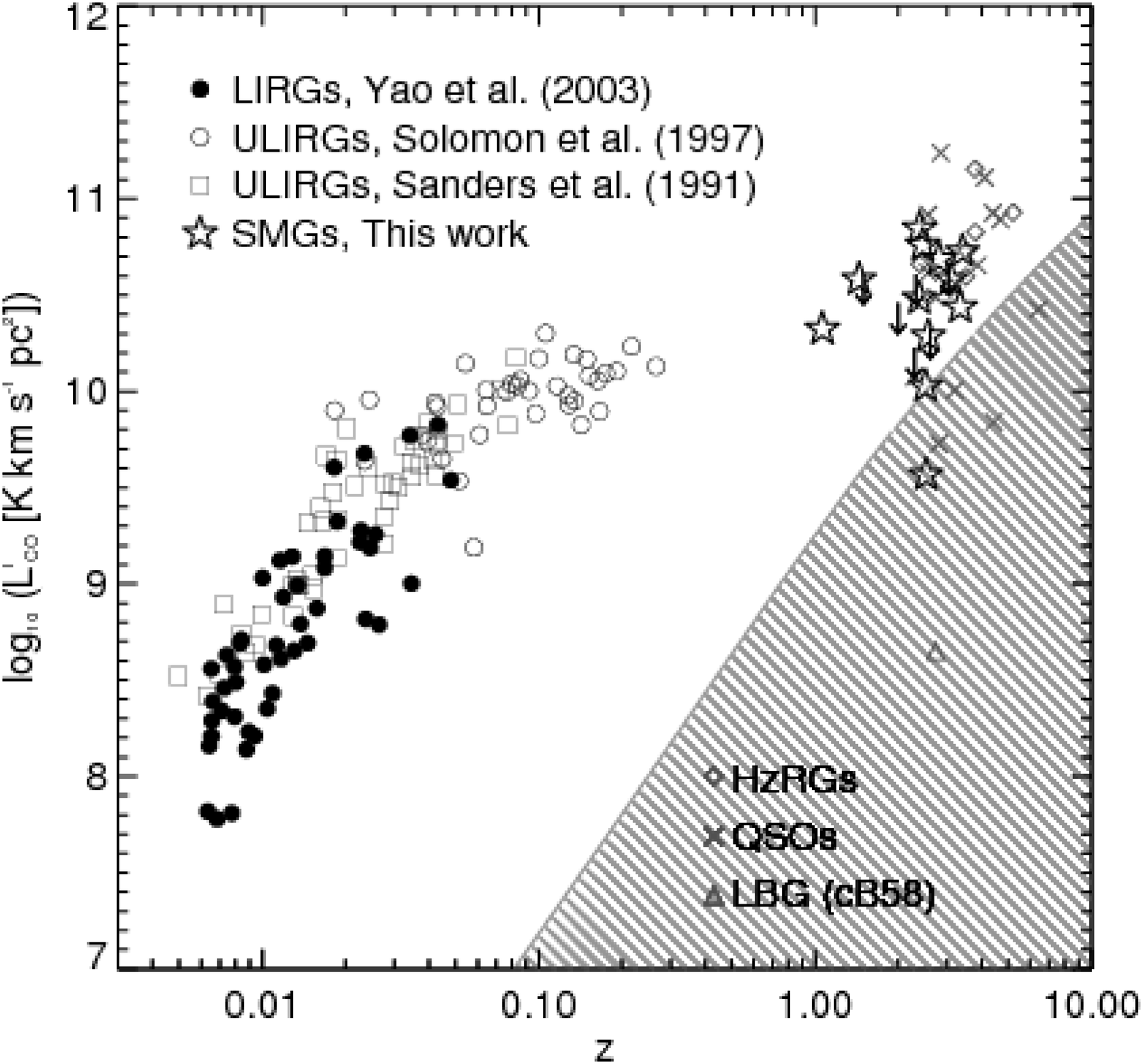}
\caption[$L'_{\mbox{\tiny{CO}}}$ versus redshift for local (U)LIRGs and
SMGs]{The relationship of $\log_{10} (L'_{\mbox{\tiny{CO}}})$ with redshift for the (U)LIRGs
samples of Sanders et al.\ (1991), Solomon et al.\ (1997), and Yao et al.\ (2003),
and for the sample of high-redshift (sub)mm-selected sources presented
in this paper. The CO luminosities have been corrected for gravitational amplification
where necessary. Also shown are the (lensing-corrected) CO luminosities for HzRGs,
QSOs  and LBGs  listed in Table \ref{table:all-co-detections}, with the exception of Q\,0957$+$561 and
MG\,0414$+$0534 which have unknown gravitational magnification factors. 
The CO luminosities for the (U)LIRGs are all based on the \COJ{1}{0} line, whereas the
luminosities of the high-redshift sources are entirely derived from
\COJJ{J+1}{J} transitions with $J+1\ge2$ (Table \ref{table:all-co-detections}). Upper limits
on the CO luminosities for the 7 SMGs not detected in CO are represented by downward pointing arrows.
The hashed area denotes the region where sources are precluded from detection
due to an integrated CO($3-2$) flux limit of $I_{\mbox{\tiny{CO}}} = 0.3$\,Jy\,km\,s$^{-1}$. Of our sample, only SMM\,J16359$+$6612
with its very large amplification factor ($\mu = 22$) is detectable below this limit.}
\label{fig:z-lco}
\end{center}
\end{figure}

In Fig.\ \ref{fig:z-lco} we plot the CO luminosities of the SMG sample as a
function of redshift along with the \COJ{1}{0} luminosities of the (U)LIRG
samples of SA91, SO97, and Y03. From Fig.\ \ref{fig:z-lco} it is clear that 
while the most luminous ULIRGs have
CO luminosities comparable to those of the faintest detected SMGs, SMGs are
generally more CO luminous than the local (U)LIRGs. In \S \ref{section:CO-H2} we
found the median CO line luminosity of the SMG sample to be
$\langle L'_{\mbox{\tiny{CO}}}\rangle = (3.8\pm 2.0)\times 10^{10}$\LCOunits, which is almost a factor
of four greater than the average CO luminosity of the ULIRGs from SO97.
Furthermore, while the CO luminosities of the (U)LIRGs are all based on the
lowest and least excitation-biased \COJ{1}{0} rotational line, all of the SMGs
are detected in higher transitions. Hence, the CO luminosities based on the
high-$J$ lines for SMGs should be regarded as lower limits on the \COJ{1}{0}
luminosities, unless the lines are fully thermalised at high temperatures and optically thick.

The (U)LIRGs lie on a well-defined locus in Fig.\ \ref{fig:z-lco}, which is
extended to the highest redshifts and CO luminosities by the SMG sample.
The lower boundary of this trend is determined by the sensitivity limits of
the CO observations, while the upper bound is a real astrophysical limit.  
The steady decline in the CO luminosity of SMGs to
ULIRGs and LIRGs seen in Fig.\ \ref{fig:z-lco} is likely to reflect the evolution in
the molecular gas-content of the most luminous starburst galaxies as a function
of redshift, despite the expected increasing mass of a typical galaxy with cosmic time.
A possible caveat is the assumption of a constant conversion factor in (U)LIRGs
and SMGs. In (U)LIRGs the \XCO\  conversion factor is about 4.5 times lower than in 
normal spiral galaxies, and it is possible that for the yet more luminous SMGs it is
even lower.

Also shown in Fig.\ \ref{fig:z-lco} are the CO luminosities of all HzRGs and
QSOs detected to date (Table \ref{table:all-co-detections}), as well as the only optically-selected
Lyman break galaxy (LBG), MS\,1512$-$cB58 ($z=2.73$), detected in CO (Baker et al.\ 2004).
Their large apparent scatter is due to many of these
objects being gravitationally-lensed, increasing the effective sensitivity
limits of their observations beyond the capabilities of current instruments in blank
fields. SMGs are seen to have CO luminosities comparable to the most luminous HzRGs
(e.g.\ Papadopoulos et al.\ 2000; De Breuck et al.\ 2003a, 2003b) and QSOs
(e.g.\ Carilli et al.\ 2002a; Walter et al.\ 2003).  
In contrast, the CO luminosity of the typical-luminosity LBG cB58, $4.2 \times 10^{9}$\LCOunits, is
nearly two orders of magnitude ($\sim 90\times$) less than the median value for
SMGs, despite its similar redshift. With only one LBG detected in CO so far it
is not possible to make any meaningful conclusions about the molecular gas content
of this population, except that the low CO luminosity of cB58 compared to
that of SMGs is consistent with the faintness of LBGs at submm wavelengths 
and their relatively low (compared to SMGs) star-formation rates as inferred from optical spectroscopy
(Blain et al.\ 1999; Adelberger \& Steidel 2000).

\subsection{CO line widths and dynamical masses}\label{section:clwadm}

In \S \ref{section:lwdm} we found that $\gs 33$ per cent of the CO-detected SMGs
have double-peaked CO profiles. This is comparable to the fraction of 
double-peaked CO spectra in local ULIRGs (SO97).
The median {\sc fwhm} CO linewidth for the SMG sample is $780\pm
320$\kms, about 3 and 4 times larger than the averages for the SA91 and SO97
ULIRG samples, respectively. Since the dynamical mass depends
on the square of the velocity dispersion ($M_{dyn} \sim R \sigma^2$), this
would naively suggest that the dynamical masses of the SMGs are $\sim 9$--$16$ times larger
than that of ULIRGs (if the gas kinematics
sample similar radii in the galaxies). From detailed interferometric studies of the circumnuclear gas in 
local ULIRGs Downes \& Solomon (1998) find a median dynamical mass of 
$\sim 6\times 10^{9}\,\Msolar$ within $R\ls 0.6$\,kpc -- about $20\times$
smaller than the mass enclosed within $R\ls 2$\,kpc 
in SMGs (see Table \ref{table:sample-average}).  Assuming an isothermal
mass distribution, the SMGs will have dynamical masses within 0.6\,kpc roughly $6\times$
that of local ULIRGs.

\begin{figure}
\begin{center}
\includegraphics[width=1.0\hsize,angle=0]{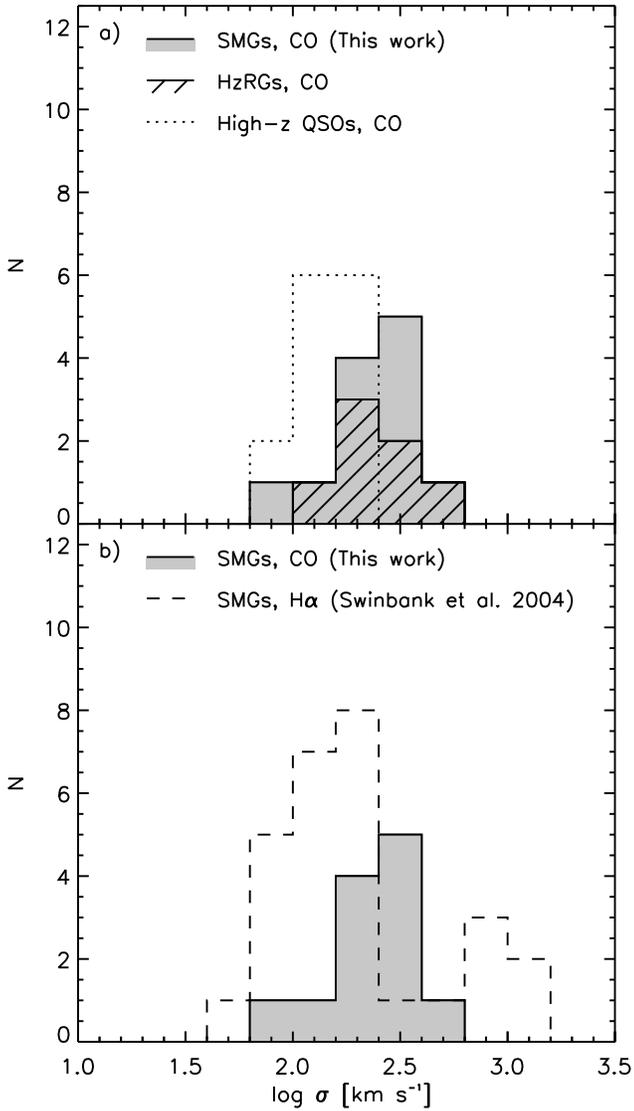}
\caption[Distribution of CO velocity dispersions for high-redshift
QSOs/HzRGs and SMGs.]{\textbf{a)} Distribution of CO velocity dispersions for 
high-redshift QSOs and HzRGs from Table 2 and our sample of
CO-detected SMGs. In the cases of the two HzRGs 4C\, 60.07 and TN\, J0924$-$2201 we
have adopted {\sc fwhm} linewidths of 1000 and 325\,\kms, respectively.  
In \textbf{b)}, we compare with the
measurements of the line widths of the H$\alpha$ line from near-infrared
spectroscopy of a large sample of SMGs (Swinbank et al.\ 2004).}
\label{fig:vdisp}
\end{center}
\end{figure}

We now turn to the comparison between SMGs and other high-redshift galaxy
populations. Fig.\ \ref{fig:vdisp}a compares the distribution of velocity
dispersions of our SMG sample with that of all high-redshift QSOs and HzRGs
detected in CO to date.
While there is a substantial overlap between the SMG and QSO distributions, the
latter is seen to peak at somewhat lower velocity dispersions than SMGs and we
note that all QSOs have $\log \sigma \le 2.4$ (corresponding to {\sc fwhm}~$
\le 590$\,\kms). Also, no QSO detected in CO show evidence of a double peaked
line profile (although a few are spatially resolved -- Papadopoulos et al.\ 2001; Carilli et al.\ 2002a). 
A Kolmogorov-Smirnov test suggests that the two
distributions are formally different at the 95 per cent confidence level. 
The apparently lower CO linewidths of QSOs is not clear, although it is
possible that they are either intrinsically lower-mass systems or have smaller gas disks.  
In contrast, there is no discernable difference between the CO line widths of the SMGs and
those HzRGs detected in CO.  This suggest that the brightest SMGs have similar
dynamical masses, and therefore possibly similar dark-matter halo masses, to
HzRGs which are believed to be amongst the most massive objects in the high-redshift Universe
(inferred from both the presence of a supermassive black hole and the strong
clustering of HzRGs -- e.g.\ Ford et al.\ 1994; Kooiman et al.\ 1995).
However, such a conclusion is uncertain due to the small number of objects
involved and will have to wait until a larger, more uniform sample of HzRGs has
been observed in CO.

%
%

In Fig.\ \ref{fig:vdisp}b we compare the CO line widths from our SMG sample with measurements
of the  width of the H$\alpha$ line from a recent near-infrared 
spectroscopic survey of 28 SMGs and high redshift, optically-faint, 
radio-selected starburst galaxies (OFRGs --- see e.g.\ Chapman et al.\ 2005) by Swinbank et al.\ (2004). 
The two distributions span a similar range of velocity dispersions, but the CO
sample is skewed towards higher values, resulting in a higher median value.
We find that the average ratio between the CO and H$\alpha$ line widths for SMGs is
$\langle${\sc fwhm}$_{\mbox{\tiny{CO}}}\rangle/\langle${\sc
fwhm}$_{\mbox{\tiny{H}}\alpha}\rangle \sim 2$.  
The main reason for this is likely to be due to the CO tracing multiple components in the SMG
whereas the long-slit H$\alpha$ spectroscopy is identifying only
a single component due to the slit orientation or due to extreme obscuration
in the second component.
Part of the discrepancy may also reflect dust-obscuration within the galaxy providing
only a partial view of the kinematics of the system, in contrast to the obscuration-independent
measurement provided by CO.
If this is the case, it could mean that the dynamical masses of SMGs
are larger than suggested from their rest-frame optical spectra.

\subsection{Star-formation efficiency and the $L'_{\mbox{\tiny{CO}}}$-$L_{\mbox{\tiny{FIR}}}$ relation}\label{section:lfir-lco} 

While the luminosity of the CO lines is an indicator of how much molecular gas a galaxy contains, it
does not necessarily measure how efficiently this gas is being turned into stars.  The
star-formation efficiency (SFE) can be estimated from the ratio of the far-infrared
luminosity of the system to the amount of
molecular gas available to form stars, i.e.\ SFE\,$ =
L_{\mbox{\tiny{FIR}}}/M(\mbox{H}_2)$.

%
%
\begin{figure}
\begin{center}
\includegraphics[width=1.0\hsize,angle=0]{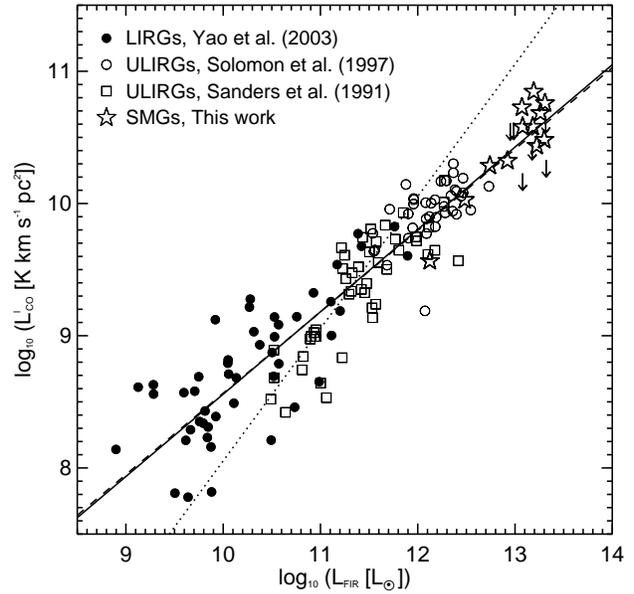}
\caption[$L_{\mbox{\tiny{CO}}}$ vs.\ $L'_{\mbox{\tiny{FIR}}}$]{
A comparison of the CO and far-infrared luminosities
($L'_{\mbox{\tiny{CO}}}$, $L_{\mbox{\tiny{FIR}}}$) for SMGs and local
ULIRGs. As in Fig.\ \ref{fig:z-lco} the arrows denote the 6 non-detections.
The solid and dashed lines represent fits of the form $\log
L'_{\mbox{\tiny{CO}}} = \alpha \log L_{\mbox{\tiny{FIR}}} + \beta$ to
all three (U)LIRGs samples and to the combined (U)LIRGs and SMGs samples, respectively.
The dotted line is the best fit to all four samples but where
the slope has been fixed to unity. }
\label{fig:lfir-lco}
\end{center}
\end{figure}

The far-infrared luminosities of the SMGs were estimated following Neri et al.\ (2003), 
$L_{\mbox{\tiny{FIR}}} = 1.9\times 10^{12}S_{850} $,
where $L_{\mbox{\tiny{FIR}}}$ is in $\Lsolar$ and $S_{850}$ in mJy.
This assumes a modified grey-body model for the 
far-infrared emission with a dust temperature of $T_d = 40$\,K and 
emissivity $\propto \nu^{1.5}$. The median far-infrared luminosity of the SMG
sample is $\langle L_{\mbox{\tiny{FIR}}}\rangle = (1.5\pm 0.7)\times 10^{13}\,\Lsolar$ which
is nearly an order of magnitude larger than the most luminous local ULIRGs (Solomon et al.\ 1997) .
Assuming that the bulk of the far-infrared luminosity is powered by 
star-formation (Frayer et al.\ 1998;
Alexander et al.\ 2005), we find a median star-formation efficiency of 
$\langle SFE\rangle = 450\pm 170\,\Lsolar\,\Msolar^{-1}$ for the SMG sample, in
agreement with the initial findings of Neri et al.\ (2003). 
This is somewhat higher than the typical starformation efficiency of the local ULIRGs studied by SO97 (
$\langle SFE\rangle = 180\pm 160 \,\Lsolar\,\Msolar^{-1}$ -- using a conversion factor of 
\XCO\,$ = 0.8$\,(K\,km\,s$^{-1}$\,pc$^2$)$^{-1} \Msolar$).

Perhaps a more straightforward measure of the star-formation efficiency is the
continuum-to-line luminosity ratio,
$L_{\mbox{\tiny{FIR}}}/L'_{\mbox{\tiny{CO}}}$, since it does not depend on
\XCO. Locally, (U)LIRGs are observed to follow a scaling relation between
$L'_{\mbox{\tiny{CO}}}$ and $L_{\mbox{\tiny{FIR}}}$ with the
more far-infrared luminous galaxies having proportionally
higher CO luminosities (Rickard \& Harvey
1984; Young et al.\ 1984, 1986; Sanders \& Mirabel 1985).  In
Fig.\ \ref{fig:lfir-lco} we have plotted the SMGs onto the
$L'_{\mbox{\tiny{CO}}}$--$L_{\mbox{\tiny{FIR}}}$ diagram along with the (U)LIRGs
from our three low-redshift comparison samples.
The SMGs extend the general trend seen for the local (U)LIRGs out to far-infrared
luminosities $\gs 10^{13}\,\Lsolar$. 
A power-law fit to all three (U)LIRG samples 
yields $\log L'_{\mbox{\tiny{CO}}} = (0.62\pm 0.09) \log L_{\mbox{\tiny{FIR}}}
+ (2.41\pm 1.06)$. Including the SMGs in the fit yields 
$\log L'_{\mbox{\tiny{CO}}} = (0.62\pm 0.08) \log L_{\mbox{\tiny{FIR}}} + (2.33\pm 0.93)$,
i.e.\ virtually no change in the fit at all. The combined SMG and local ULIRG samples have a 
statistically-significant correlation, as shown by a
Spearman's rank-order correlation test, which yields a probability of
$P<0.0001$ that a random (uncorrelated) data set could result in the
observed correlation coefficient ($r_S = 0.93$).  
From Fig.\ \ref{fig:lfir-lco} there is
even some evidence of a $L'_{\mbox{\tiny{CO}}}$--$L_{\mbox{\tiny{FIR}}}$ correlation 
within the SMG population itself (the probability of obtaining the observed correlation, 
$r_S=0.70$, by chance is $P<0.0142$). 
Moreover, the slope of the 
$L'_{\mbox{\tiny{CO}}}-L_{\mbox{\tiny{FIR}}}$ correlation inferred from the fit
is significantly less than unity. This is clearly illustrated in 
Fig.\ \ref{fig:lfir-lco} where the locus defined by the data points is 
seen to have a shallower slope than the line of equality.

\begin{figure} 
\begin{center}
\includegraphics[width=1.0\hsize,angle=0]{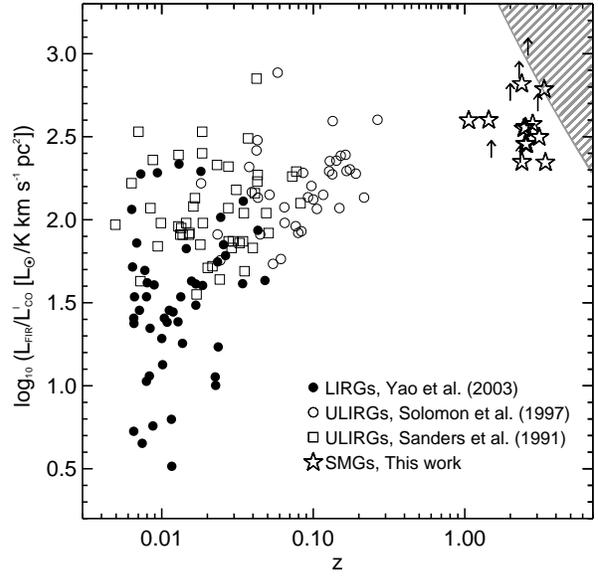}
\caption[The star-formation efficiency as indicated
$L_{\mbox{\tiny{FIR}}}/L'_{\mbox{\tiny{CO}}}$ versus redshift.]  {The
star-formation efficiency as indicated by the observed ratio 
$L_{\mbox{\tiny{FIR}}}/L'_{\mbox{\tiny{CO}}}$ as a function of redshifts. Upward pointing
arrows represent the 6 SMGs not detected in CO. The hashed region represents
the no-detection zone for sources with integrated CO($3-2$) fluxes of 
$I_{\mbox{\tiny{CO}}}\le 0.3$\,Jy\,km\,s$^{-1}$ and submm fluxes of $S_{850\mu\mbox{\tiny{m}}} \le 4$\,mJy.}
\label{fig:z-lfirlco}
\end{center}
\end{figure}

The observed slope of the
correlation in Fig.\ \ref{fig:lfir-lco} implies that the $L_{\mbox{\tiny{FIR}}}/L'_{\mbox{\tiny{CO}}}$
ratio increases with $L_{\mbox{\tiny{FIR}}}$. 
Several studies have already shown that ULIRGs have
higher $L_{\mbox{\tiny{FIR}}}/L'_{\mbox{\tiny{CO}}}$ ratios than the less
luminous LIRGs, which in turn have higher ratios than spiral galaxies and GMCs
(e.g.\ Sanders et al.\ 1986; Solomon \& Sage 1988). 
This trend is shown in 
Fig.\ \ref{fig:z-lfirlco} where $L_{\mbox{\tiny{FIR}}}/L'_{\mbox{\tiny{CO}}}$ 
has been plotted as a function of redshift for the LIRG, ULIRG and SMG samples.
The median $L_{\mbox{\tiny{FIR}}}/L'_{\mbox{\tiny{CO}}}$ 
ratios for LIRGs (Y03) and ULIRGs (SO97) are $50\pm 30$ and $160\pm 130\,\Lsolar\,(\mbox{K\,km\,s}^{-1}\mbox{pc}^2)^{-1}$, 
respectively.  The median far-infrared  to CO luminosity ratio for the SMGs is 
$360 \pm 140\,\Lsolar\,(\mbox{K\,km\,s}^{-1}\mbox{pc}^2)^{-1}$, which
is about a factor of two higher than the median value for ULIRGs, and
suggests that although some overlap exists between the two populations,
in general SMGs have higher star-formation efficiencies, or a larger contribution
to their far-infrared luminosities from a source other than star formation, e.g.\ an AGN.
Although AGN are frequently found in SMGs using extremely deep
X-ray observations, these observations also suggest that the AGN are bolometrically
insignificant in the vast majority of SMGs (Alexander et al.\ 2003, 2005).

%
%

We underline that the above findings are strongly biased towards luminous SMGs,
as they are based on a sample of predominantly bright SMGs,
This selection effect in conjunction with the CO detection limit of our survey
prevent us from probing the lower right portion of Fig.\ \ref{fig:z-lfirlco}, 
where objects with low $L_{\mbox{\tiny{FIR}}}/L'_{\mbox{\tiny{CO}}}$ ratios
reside. It is possible that such objects, which are found locally (see Fig.\ \ref{fig:z-lfirlco}), 
will also be uncovered at high redshifts by future 
deep submm and CO observations of a complete sample of high-$z$ starburst galaxies.  
Intriguingly, however, the $L_{\mbox{\tiny{FIR}}}/L'_{\mbox{\tiny{CO}}}$ ratio of 
SMM\,J16359$+$6612, which has a far-infrared and CO luminosity
comparable to the least luminous ULIRGs in the SO97 sample, is higher
than almost all the sources in that sample ($360\,\Lsolar\,(\mbox{K\,km\,s}^{-1}\mbox{pc}^2)^{-1}$).

As a final check we repeated the above analysis using SMG far-infrared
luminosities derived from their radio fluxes and using the radio-FIR correlation
(Condon 1992; Yun, Reddy \& Condon 2001). There is tentative evidence that this correlation applies out to
high redshifts (Garrett 2002; Appleton et al.\ 2004), and arguably provides
a more reliable estimate of the far-infrared luminosity than that based on
the submm flux which has significant uncertainty associated with it due to the
unknown dust properties of SMGs. The far-infrared luminosities derived 
using the radio information are generally
larger, and the $L'_{\mbox{\tiny{CO}}}-L_{\mbox{\tiny{FIR}}}$ correlation 
therefore shallower ($\alpha = 0.59\pm 0.08$). In this case, the SMG sample also
exhibits a significant correlation, although this can
partly be explained by the luminosity distance-squared stretching of the data points.

\subsection{Discussion}\label{section:discussion}

In the previous three sub-sections it was found that SMGs have on average four times larger CO luminosities,
3--4 times larger line widths and a median $L_{\mbox{\tiny{FIR}}}/L'_{\mbox{\tiny{CO}}}$ ratio 
twice that of local ULIRGs.
As argued in \S \ref{section:lwdm}, the large line widths and molecular
gas masses seen in SMGs are difficult to reconcile with a scenario 
in which the gas resides in a circumnuclear disk. This is in contrast to what is 
observed in local ULIRGs where the bulk of the gas is found in a $R\sim 0.5$\,kpc disk or
ring.  In SMGs the gas is more likely to be distributed on scales of 2--3\,kpc, as suggested
by their radio morphologies (Chapman et al.\ 2004) as well as by high-resolution
CO observations of SMGs (Genzel et al.\ 2003; Tacconi et al.\ 2005).
The typical gas-to-dynamical mass
fraction in SMGs was estimated in \S \ref{section:lwdm} to be $\sim 0.3$ -- almost two times
higher than the median gas mass fraction in ULIRGs 
($\sim 0.16$, Downes \& Solomon 1998), assuming that the CO--H$_2$ conversion factor 
is the same for ULIRGs and SMGs. The local gas mass fraction in ULIRGs is measured within the
molecular disk ($R\ls 0.5$\,kpc) -- but when averaged over scales of 2-3\,kpc, which are the typical
scales probed by our CO observations of SMGs, the gas fraction is likely to be significantly smaller.
Although, we caution that some of the above
arguments are based on a relatively small number of sources detected at low signal-to-noise
or have non-negligible uncertainties associated with them (such as the CO conversion factor), 
they do seem to suggest, together with the generally higher far-infrared luminosities of SMGs,
that the latter are neither high-redshift replicas of
local ULIRGs, nor simply scaled-up versions. Rather SMGs appear to more gas-rich and more
efficiently star-forming than local ULIRGs.

In this respect it is important to draw attention to observational studies of the depletion
of molecular gas with starbursts in ULIRGs.  In a sample of more than 50
(U)LIRGs, Gao et al.\ (1999) found a correlation between the \COJ{1}{0}
luminosity and the projected separation of the merging nuclei, which they took
as evidence for molecular gas being rapidly depleted due to intense star
formation as the merger progresses. If a similar picture applies at high
redshift, the large CO luminosities and gas masses we find for SMGs would imply that these
systems are extended, and caught in the early stages of merging. 
However, this might be a premature conclusion since Rigopoulou et al.\ (1999) did not
find any evidence for a correlation between the gas mass and merger phase for local ULIRGs.

If the above trends are verified by future observations, it is interesting to speculate
what physical mechanisms could be responsible for the differences between local ULIRGs
and SMGs?  Both populations appear to result from mergers and strong interactions
(e.g.\ Sanders et al.\ 1991; Chapman et al.\ 2003c), so what makes the starbursts so much more
efficient in SMGs?
Numerical simulations have compared the gas flow in major mergers between two
galaxies with strong bulges, and between two, gas-rich but bulge-less galaxies 
(Barnes \& Hernquist 1996; Iono, Yun \& Mihos 2004). 
In the first case, the gas forms shocked, dense filaments
which dissipate energy and flow towards the centre where
they form a nuclear ring which is stabilised by the gravitational potential of
the bulge -- slowing the rate of star formation. 
In the second scenario, the merger between two gas-rich but bulge-less 
galaxies, the gas forms an extended ($\sim 6$\,kpc), gravitationally-significant
bar-like structure 
which allows the gas to funnel towards the centre.  However, without
the stabilising
influence of the bulges in the progenitor galaxies, the gas cannot form a
stable ring or disk and its density is further increased through the formation
of bar-like structures, leading to a wide-spread and vigorous starburst.

These two theoretical scenarios suggest a structural difference between the
progenitor galaxies may be at the heart of the differences in behaviour
between ULIRGs and SMGs.  The suggested structural differences are consistent with
the expectations of the typical galaxies involved in mergers at
low- and high-redshifts, with the former involving disk-galaxies
with significant bulge components  (e.g.\ Lilly et al.\ 1998),
while the progenitor galaxies of SMGs at high redshift are much more
likely to be gas-rich, disk-dominated systems with little or no
bulge component (Wyse, Gilmore \& Franx 1997; Ravindranath et al.\ 2004).
If such structural differences between local and distant starbursts are real, 
the next question is: at what redshift does the transition occur?
In \S \ref{section:CO-H2}  we found that SMGs at $z\sim 3$ have a higher CO detection
rate than SMGs at $z\sim 2$ -- consistent with the former being more 
gas-rich. Secondly, the slight preponderance, albeit tentative, of double-peaked CO line
profiles at $z\sim 2$ is consistent with that of local ULIRGs and
in contrast to SMGs at $z\sim 3$
(see \S \ref{section:CO-H2} and \S \ref{section:lwdm}). 
Although, these trends are tentative and need a larger survey
to confirm or disprove them, together they
may indicate a difference in the physical processes responsible for
triggering intense star formation
in massive galaxies at $z\sim 2$ compared to $z\sim 3$, with major mergers being responsible
for much of this activity at lower redshifts (as is the case for local ULIRGs), but a 
separate mode (requiring less intense perturbations) capable of triggering similar bursts
in the more gas-rich systems present at even higher redshifts.

\bigskip

In \S \ref{section:lfir-lco} we showed that luminous SMGs extend the 
$L_{\mbox{\tiny{CO}}}-L'_{\mbox{\tiny{FIR}}}$ relation of
local ULIRGs to higher redshifts and luminosities, and accordingly have higher
star-formation efficiencies than ULIRGs.
Although, the star-formation efficiencies found for the SMGs could be
severely overestimated if an AGN contributes significantly to the far-infrared
luminosity, the detection of large amounts of molecular gas 
in SMGs along with recent X-ray (Alexander et al.\ 2003, 2005) and radio
studies (Chapman et al.\ 2004)
strongly suggest that the bulk of the far-infrared  emission 
from SMGs is powered by a large-scale starburst and
not from an AGN.  While CO is a good indicator of the total
metal-rich H$_2$ gas reservoir, it may be a worse indicator of the amount of dense gas
present ($n \ge 10^5\,\mbox{cm}^{-3}$), that actually fuels star formation
(Carilli et al.\ 2004).  The latter
could be particularly true in the tidally disrupted giant molecular clouds (GMCs)
expected in ULIRGs, where a diffuse phase may dominate the CO emission
but has little to do with star formation (Downes \& Solomon 1998; Sakamoto et al.\ 1999). 
Such a diffuse phase could be even
more pronounced in SMGs with their more extended distributions
(Chapman et al.\ 2003c; Smail et al.\ 2004).  This would explain why the
$L_{\mbox{\tiny{FIR}}}/L'_{\mbox{\tiny{CO}}}$ ratio is found to be such a
strong function of $L_{\mbox{\tiny{FIR}}}$, increasing for merging systems
usually associated  with the highest far-infrared  luminosities. Interestingly, recent work
shows that the SFE of dense gas, parametrised by the
$L_{\mbox{\tiny{FIR}}}/L_{\mbox{\tiny{HCN(1-0)}}}$ ratio (the HCN \J{1}{0}
critical density is $2 \times 10^5\,\mbox{cm}^{-3}$), remains constant from
GMCs all the way to ULIRG system (Gao \& Solomon 2003; Solomon et al.\ 2003; Carilli et al.\ 2004).

\bigskip

The median star-formation rate of the 18 SMGs is $\langle SFR \rangle \simeq 700\,\Msolar\,\mbox{yr}^{-1}$, where we have assumed
a Salpeter IMF and a very conservative limit corresponding to the
starburst contributing only 50 per cent of
the far-infrared luminosity (see Omont et al.\ 2001, but c.f.\ Alexander et al.\ 2005).
Our findings in \S \ref{section:clagm} and \S \ref{section:clwadm} have shown 
that SMGs are massive galaxies with enough molecular gas in them
to sustain such a large star-formation rate for 
$\tau_{SMG} \sim M(\mbox{H}_2)/SFR \sim 3\times 10^{10}\,\Msolar/ 700\,\Msolar\,\mbox{yr}^{-1} \sim
40\,\mbox{Myr}$. 
By the end of such a burst most of the stellar mass 
corresponding to that of a massive spheroid would be in place.
A gas depletion timescale of $\tau_{SMG}\sim 40$\,Myr is comparable 
to the typical starburst ages ($\sim 100$--$200$\,Myr) derived from photometric modeling of the broad band
optical/near-infrared colours of SMGs (Smail et al.\ 2004), although we note that 
large uncertainties are associated with both methods. 
For example, the CO observations only probe the gas within the central 10\,kpc, and
as a result the neutral gas --  which is likely to be distributed on larger scales -- has not been 
included in our estimate of the gas consumption time scale.
If the H\,{\sc i} gas is brought in from $\gs 10$\,kpc radius at 
$\sim 200$\,km\,s$^{-1}$ it would reach the central regions in less than
$50$\,Myrs, where it could help sustain the vigorous starformation.
Furthermore, the above gas consumption time scale assumes a continuous starburst
until all the gas is used -- an unlikely scenario since it ignores the negative feedback 
effects from new-born massive stars and supernovae.
For example, one could imagine the starburst terminating prematurely if the gas is removed
by starburst- and/or AGN-driven winds only to fall back
onto the galaxy at a later stage to fuel a second starburst, 
thus making the true gas exhaustion time scale longer. 
The 40\,Myr should therefore be considered a strict lower limit on the
starburst phase of SMGs.

An alternative estimate of the duration of the SMG phase can be made from the
recent findings by Page et al.\ (2004) that the 15 per cent of QSOs at
$z\sim 2$ which show absorption in their X-ray spectrum are detectable in the submm.
Since only 3 per cent of radio-identified SMGs are QSOs
(Chapman et al.\ 2005) this means that SMGs would have a typical lifetime which
is $\sim 0.15/0.03 = 5$ times longer than that of QSOs. Adopting a QSO lifetime
of $40\,\mbox{Myr}$ (Martini \& Weinberg 2001) yields a SMG life expectancy of
$\tau_{SMG} \sim 200\,\mbox{Myr}$ --- again larger than the gas consumption time scale.

If we assume that a large fraction of the gas mass of SMGs is 
eventually converted into
stars (perhaps through repeated cycles of
expulsion, infall and star formation), what are the resulting stellar masses of the descendents? Clearly
SMGs already contain (perhaps substantial) stellar populations (Smail et al.\ 2004), however
the stellar masses for SMGs are difficult to estimate due to significant dust extinction
and the resulting uncertainties from the degeneracies  between dust reddening and age
in evolutionary 
spectral synthesis models. The best estimates to date, based on 
$IJK$ photometry of 96 SMG and optically-faint $\mu$Jy
radio galaxies (Smail et al.\ 2004), suggest typical
stellar masses of $M_{stars}\simeq 3\times 10^{10}\,\Msolar$.
Thus, combining the gas and stellar mass estimates we find that on average SMGs have
baryonic masses of $\gs 6\times 10^{10}\,\Msolar$, which is comparable to
the masses of local early type $L^*$ galaxies.

The discussion above suggests that SMGs are in fact the progenitors of
massive spheroids in the present-day Universe and that the build-up
of the stellar population occurs rapidly ($40 \ls \tau_{SMG} \ls 100$\,Myr), consistent with
the old homogeneous stellar populations of local ellipticals. Further evidence in support of
this is a) the strong clustering claimed for SMGs (Blain et al.\ 2004b),
and b) the fact that the rest-frame optical properties of SMGs match the bright end of the luminosity
function of spheroidal galaxies in nearby clusters (Smail et al.\ 2004).

\section{Implications for structure-formation models}\label{section:implications}

Semi-analytical models of galaxy formation and evolution (Cole et al.\ 1994;
Kauffmann et al.\ 1999; Cole et al.\ 2000; Somerville et al.\ 2001) have enjoyed a
fair degree of success in reproducing the properties
of galaxies in the local Universe, e.g.\ the luminosity
function, the distribution of colours and disk scale lengths of galaxies, the
observed mix of morphologies and the Tully-Fisher relation (e.g.\ Kauffmann et
al.\ 1999; Cole et al.\ 2000).  However, the key observable
with which to benchmark models of galaxy formation and evolution is the
mass assembly of galaxies, and in particular the assembly of baryonic mass, as
a function of redshift.

Genzel et al.\ (2003) pointed out that if the extremely large baryonic mass ($\gs 10^{11}\,\Msolar$) 
of SMM\,J02399$-$0136 was representative of the bright SMG population, then the 
high surface density of such sources would imply that the abundance of very massive
baryonic systems at high redshift was about an order of magnitude larger than
predicted by semi-analytical models (see also Tecza et al.\ 2004).

%
%
\begin{figure}
\begin{center}
	\includegraphics[width=1.0\hsize,angle=0]{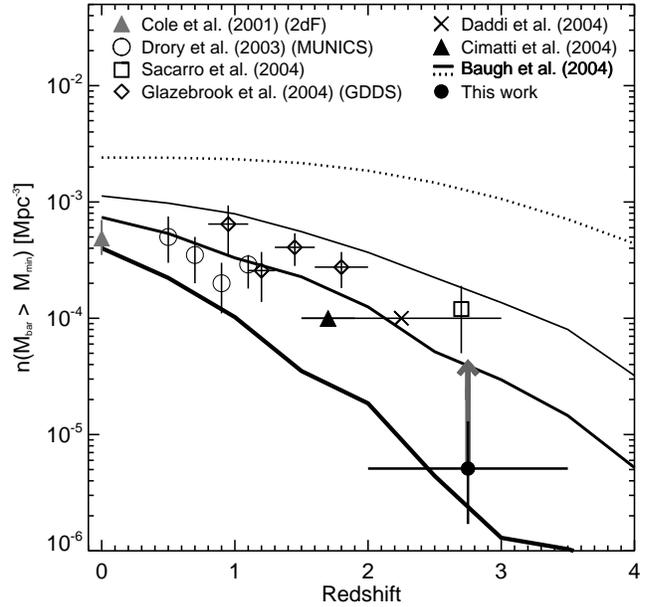}
\caption[Co-moving number density of galaxies with baryonic masses 
$\sim 6\times 10^{10}\,\Msolar$ as derived from CO observations of SMGs (filled circles).]
{The co-moving number density of galaxies with baryonic masses $\sim 6\times 10^{10}\,\Msolar$ 
as derived from CO observations of SMGs. The
GALFORM model (Cole et al.\ 2001; Baugh et al.\ 2004) predictions
of the abundances of galaxies with baryonic masses $\ge 5\times$, $7\times$, 
and $10\times 10^{10}\,\Msolar$ as a function of redshift are shown
as thin, medium and thick solid lines, respectively.
The dotted curve represents the total baryonic matter content available 
in $\ge 10^{11}\,\Msolar$ dark matter halos, and is obtained
by scaling the abundance of halos with the cosmological baryon-to-dark matter
density ($\Omega_b/\Omega_{DM} = 0.13$). This provides a 
strict upper limit on the number density of massive baryonic
galaxies at a given redshift. The observations can be reconciled within
the CDM framework, provided that $\sim 10$ per cent of all baryons within dark matter halos are
rapidly assembled into galaxies.}  
\label{fig:mass-function}
\end{center}
\end{figure}

Here we repeat this analysis using the 10 SMGs in our sample which lie in the redshift
range $z=2$--$3.5$ and have reliable gas mass estimates, to  
put lower limits on the co-moving number 
density of massive galaxies at high redshifts.
The redshift interval considered translates into $\sim 1.5$\,Gyr in terms of elapsed cosmic
time, and corresponds to a co-moving volume of $1.8\times 10^{7}$\,Mpc$^{3}$ per square degree.
The average submm flux of the 10 SMGs is $\langle S_{850\mu\mbox{\tiny{m}}}\rangle = 8\pm
4$\,mJy.  The surface density on the sky of SMGs with fluxes $\gs 8$\,mJy
is $285^{+231}_{-149}$\,deg$^{-2}$ (Borys et al.\ 2003). Taking into account that a) 
about 60 per cent of the bright SMG population lie within $z=2$--$3.5$ (Chapman et al.\ 2003b,
Chapman et al.\ 2005), and b) the CO detection fraction of SMGs in the redshift range 
$2\le z \le 3.5$ is 10/14 ($\sim 71$ per cent), 
we estimate that the co-moving number density of 
galaxies with baryonic masses $\gs 6\times 10^{10}\,\Msolar$ in the redshift interval
$z=2$--$3.5$ is $\sim 5.1^{+7.6}_{-3.4}\times 10^{-6}\,\mbox{Mpc}^{-3}$.
The errors are estimated by propagating the 1-$\sigma$ limits on the submm flux
and number counts through the same calculation.

Since we only observe SMGs during their submm-luminous phase, we have to correct the 
derived space density by a factor corresponding to the ratio between the
1.5\,Gyr which has elapsed over the redshift range $z=2.0$--$3.5$ and
the typical duration of the 'SMG phase'. The latter is uncertain, but as we saw
in \S \ref{section:discussion} it is likely to lie in the range $40$--$200$\,Myr.
Adopting 200\,Myr as as a conservative upper value for the submm luminous phase,
we estimate a correction factor of $\sim 8$. 

In Fig.\ \ref{fig:mass-function} we have plotted our estimate of the co-moving
number density of $\gs 6\times 10^{10}\,\Msolar$ systems at $z\sim 2.8$ 
and the (conservative) correction factor which must be applied.
Our estimate of the abundances of massive
galaxies at $z \sim 2.8$ is in good agreement with independent
measurements at similar redshifts (Daddi et al.\ 2004; Saracco et al.\ 2004),
and confirms the slow decline in the space density
of massive galaxies as a function of redshift (Genzel et al.\ 2003; Glazebrook et al.\ 2004).

The abundance tracks of $\ge 5\times 10^{10}$ and $\ge 7\times 10^{10}\,\Msolar$ systems
as predicted by the most recent GALFORM models (Cole et al.\ 2001; Baugh et al.\ 2004) 
are seen to envelope our estimated volume density.
There are two reasons for the closer agreement between the observations and models than previous work: a) the typical
baryonic mass of bright SMGs is not $2$--$3\,m^*$ but rather $\sim 0.6\,m^*$ and
b) the new GALFORM model, which employs a top-heavy initial mass function (IMF)
in order to account for the 850-$\mu$m number counts (Baugh et al.\ 2004), also predicts
higher abundances of massive galaxies at high redshifts than previous models. Other semi-analytical models
(Fontana et al.\ 2004; Granato et al.\ 2004) have also modified their recipes so that their
models fit the data. 

We stress, however, that by using a SMG timescale of $200$\,Myr, and not
$40$\,Myr as suggested by the CO observations, our
estimate of the correction factor is a conservative one.
Furthermore, if the population of OFRGs,
whose faint levels of submm flux is believed to be due their higher dust temperatures
(Blain et al.\ 2004a; Chapman et al.\ 2005), have similar gas masses as classical
SMGs (Swinbank et al.\ 2004; Smail et al.\ 2004), this
would double our current estimate of the abundance of
massive baryonic galaxies at high redshifts.

We conclude that given the small number of sources and crude redshift bins used in 
our calculations, and the large uncertainty on the correction factors which have to be applied, 
there is no evidence for a severe discrepancy between observations and the latest theoretical
predictions. An increase in our sample size would not only allow us to confirm this but
also allow us to split the sample into mass bins and compare with different mass track predictions.

\section{Conclusions}

We present results from a PdBI CO Survey of SMGs with radio
counterparts. We successfully detect CO in 4 out of 10 SMGs observed, which
brings the total number of SMGs detected by our survey to 7 out of 13 observed. In
addition, we have also presented a detection of CO
based on archival observations of SMM\,J02396$-$0134.
Combining these 14 sources with four CO-detected
SMGs from the literature we have compiled a sample of 18 SMGs observed in CO
of which 12 are detected.
We use this unique sample to derive the bulk gas properties and masses of the most luminous SMGs.

\begin{itemize}

\item We find that the SMGs in our sample have a median CO luminosity of $\langle \LCO \rangle
= (3.8 \pm 2.0)\times 10^{10}$\,\LCOunits.  This corresponds to a molecular gas
mass of $\langle M(\mbox{H}_2)\rangle = (3.0\pm 1.6)\times 10^{10}\,\Msolar$ (within
$R\ls 2$\,kpc), assuming a
conversion factor of \XCO$=0.8$\,(K\,km\,s$^{-1}$\,pc$^2$)$^{-1} \Msolar$ and
optically thick, thermalised line ratios.  Although considerable uncertainty is
associated with \XCO\ it is clear that bright SMGs are amongst the most gas-rich
systems in the Universe. Comparing with local ULIRGs, we find that SMGs have
molecular gas reservoirs on average about four times greater than even the most
CO-luminous ULIRGs.  We argue that this is largely due to the evolution in the
molecular gas content of the most far-infrared  luminous galaxies with redshift.

\item In general, the SMGs in our sample have extremely broad line
profiles. The median {\sc fwhm} is $780 \pm 320$\,\kms, which is $\gs 3$ times
larger than the average \COJ{1}{0} line width of local ULIRGs. The large line
widths, and in some cases multiple-peaked CO spectra,
together with the vast amounts of molecular gas, suggests that the
brightest SMGs are merger events. We argue that the observed gas
properties are difficult to reconcile with the scenario witnessed in local
ULIRGs where the bulk of the molecular gas resides in a compact circumnuclear
disk.

\item We find that the median dynamical mass of the SMG sample is $\langle M_{dyn}\rangle = (1.2\pm
1.5)\times 10^{11}\,\Msolar$ within the central $\sim$4\,kpc. This suggests
that the brightest SMGs are amongst the most massive galaxies in the distant
Universe, comparable in mass to the most extreme HzRGs and QSOs, rather than
being high-z replicas of local ULIRGs.  Taking into account 
the stellar mass component we estimate that the total baryonic mass 
content of SMGs is $\gs 6\times 10^{10}\,\Msolar$. Thus
we conclude that not only are SMGs very massive baryonic systems, but the
baryons can account for a substantial fraction of the total mass in the central
regions.

\item We have shown that the SMGs exhibit a non-linear correlation between
far-infrared  and CO luminosity, similar to that observed for local ULIRGs. This not
only extends the $L'_{\mbox{\tiny{CO}}}$--$L_{\mbox{\tiny{FIR}}}$ relation to
higher luminosities ($L_{\mbox{\tiny{FIR}}} \sim 10^{12-14}\,\Lsolar$) but also
shows that it holds at the highest redshifts.  The main implication of this is
that SMGs have higher $L_{\mbox{\tiny{FIR}}}/L'_{\mbox{\tiny{CO}}}$ ratios than
ULIRGs, and therefore possibly higher star-formation efficiencies. However, a
clearer picture of the star-formation efficiency of SMGs has to await a future
systematic survey for high-redshift HCN, which will trace the dense
star-forming gas.

\item From the inferred molecular gas masses we estimate a typical gas
consumption timescale of $\gs 40$\,Myr. Given the large uncertanties involved, an 
SMG phase of $\sim 40$\,Myr is roughly consistent with the best age 
estimates of the starbursts in SMGs (Tecza et al.\ 2004; Smail et al.\ 2003, 2004),
which in turn agree favourably with independent estimates of the duration of the SMG phase obtained
by requiring a scenario in which SMGs last $100-200$\,Myrs before going through a $\sim10$\,Myr QSO phase at
$z\sim 2$, becoming massive evolved galaxies at $z\sim 1$, followed by a
period of passive evolution which sees them ending up as old $\gs L^\ast$
ellipticals in the present day. Furthermore, the slightly short gas consumption time scales can easily
be prolonged by invoking negative feedback processes, and may be indicative of
such processes playing an important role in the build-up of massive galaxies.

\item From our observations, we place a lower limit on the co-moving number
density of massive baryonic systems in the redshift range $z=2$--$3.5$ of
$n(M_{bar} \ge 6\times 10^{10}\,\Msolar) \gs 5.1\times 10^{-6}\,\mbox{Mpc}^{-3}$ in agreement
with results from recent near-infrared/spectroscopic surveys (e.g.\ Saracco et al.\ 2004; 
Glazebrook et al.\ 2004). Given the substantial uncertainties involved, we find no significant 
discrepancy between the data and the predicted abundances
of massive galaxies at high redshifts. 
\end{itemize}

\section*{Acknowledgments}
TRG acknowledges support from the Danish Research Council and from the
European Union RTN network, POE. IRS acknowledges support from the
Royal Society. AWB acknowledges support from NSF grant AST-0205937, from the Research 
Corporation and the Alfred P.\ Sloan Foundation. JPK thanks Caltech and CNRS for
support. We are grateful to Genevieve Soucail and
Paola Andreani for letting us use the PdBI CO data for SMM\,J02396$-$0134 and ERO\,J16540$+$4626, respectively. 
We also thank Cedric Lacey for providing the GALFORM model predictions. Finally, we thank Dennis Downes for
useful comments and suggestions and for allowing us to reproduce the PdBI CO data for SMM\,J14011$+$0252.

\bsp

\label{lastpage}


\begin{thebibliography}{99}
\bibitem[\protect\citeauthoryear{Aalto}{1995}]{Aalto-et-al-1995} Aalto S., Booth R.\ S., Black J.\ H., Johansson L.\ E.\ B., 1995, A\&A, 300, 369.
\bibitem[\protect\citeauthoryear{Adelberger}{2000}]{Adelberger-et-al-2000} Adelberger K.\ L., Steidel C.\ C., 2000, ApJ, 544, 218. 
\bibitem[\protect\citeauthoryear{Alloin}{2000}]{Alloin-et-al-2000} Alloin D., Barvainis R., Guilloteau S., 2000, ApJ, 528, L81. 
\bibitem[\protect\citeauthoryear{Appleton}{2004}]{Appleton-et-al-2004}  Appleton P.\ N.,  Fadda S.\ T., Marleau F.\ R., 2004, ApJS Spitzer Special Issue, 154, 147.
\bibitem[\protect\citeauthoryear{Alexander}{2003}]{Alexander-et-al-2003} Alexander D.\ M.\ et al., 2003, AJ, 126, 539.
\bibitem[\protect\citeauthoryear{Alexander}{2005}]{Alexander-et-al-2005} Alexander D.\ M., Bauer F.\ E.,  Chapman S.\ C., Smail I., Blain A.\ W., Brandt W.\ N., Ivison R.\ J., 2005, to appear in the  Proceedings of the ESO/USM/MPE Workshop on "Multiwavelength Mapping of Galaxy  Formation and Evolution", eds. R. Bender and A. Renzini, Venice, Italy. (astro-ph/0401129).
\bibitem[\protect\citeauthoryear{Andreani}{1995}]{Andreani-et-al-1995} Andreani P., Casoli F., Gerin M., 1995, A\&A, 300, 43.
\bibitem[\protect\citeauthoryear{Andreani}{2000}]{Andreani-et-al-2000} Andreani P., Cimatti A., Loinard L., R\"{o}ttgering H., 2000, A\&A, 354, L1.
\bibitem[\protect\citeauthoryear{Baker}{2004}]{Baker-et-al-2004} Baker A.\ J., Tacconi L.\ J., Genzel R., Lehnert M.\ D., Lutz D., 2004, ApJ, 604, 125.
\bibitem[\protect\citeauthoryear{Barger}{1999}]{Barger-et-al-1999} Barger A.\ J., Cowie L.\ L., Sanders D.\ B., 1999, ApJ, 518, L5.
\bibitem[\protect\citeauthoryear{Barnes}{1996}]{Barnes-and-Hernquist-1996} Barnes J.\ E., Hernquist L., 1996, ApJ, 471, 115.
\bibitem[\protect\citeauthoryear{Barvainis}{1994}]{Barvainis-et-al-1994} Barvainis R., Tacconi L., Antonucci R., Alloin D., Coleman P., 1994, Nature, 371, 586.
\bibitem[\protect\citeauthoryear{Barvainis}{1997}]{Barvainis-et-al-1997} Barvainis R., Maloney P., Antonucci R., Alloin D., 1997, ApJ, 484, 695.
\bibitem[\protect\citeauthoryear{Barvainis}{1998}]{Barvainis-et-al-1998} Barvainis R., Alloin D., Guilloteau S., Antonucci R., 1998, ApJ, 492, L13.
\bibitem[\protect\citeauthoryear{Barvainis}{2002}]{Barvainis-et-al-2002} Barvainis R.,  Alloin D., Bremer M., 2002, A\&A, 385, 399.
\bibitem[\protect\citeauthoryear{Baugh}{2004}]{Baugh-et-al-2004} Baugh C.\ M, Lacey C.\ G., Frenk C.\ S., Granato G.\ L., Silva L., Bressan A., Benson A.\ J., Cole S., 2004, MNRAS, in press, (astro-ph/0406069).
\bibitem[\protect\citeauthoryear{Beelen}{2004}]{Beelen-et-al-2004} Beelen A.\ et al., 2004, A\&A,423, 441.
\bibitem[\protect\citeauthoryear{Bertoldi}{2000}]{Bertoldi-et-al-2000} Bertoldi F., Menten K.\ M., Kreysa E., Carilli C.\ L., Owen F., 2000, 24th meeting of the IAU, Joint Discussion 9, Manchester, England.
\bibitem[\protect\citeauthoryear{Bertoldi}{2003}]{Bertoldi-et-al-2003} Bertoldi, F.\ et al., 2003, A\&A, 409, L47.
\bibitem[\protect\citeauthoryear{Binney \& Tremaine}{1987}]{Binney-and-Tremaine-1987} Binney J., Tremain S., 1987, "Galactic Dynamics", Princeton Series on Astrophysics, Princeton University Press.
\bibitem[\protect\citeauthoryear{Blain}{1999}]{Blain-et-al-1999} Blain A.\ W., Jameson A., Smail I., Longair M.\ S., Kneib J.-P., Ivison R.\ J., 1999, MNRAS, 309, 715. 
\bibitem[\protect\citeauthoryear{Blain}{2002}]{Blain-et-al-2002} Blain A.\ W., Smail I., Ivison R.\ J., Kneib J.-P., Frayer D.\ T., 2002, PhR, 369, 111.
\bibitem[\protect\citeauthoryear{Blain}{2004a}]{Blain-et-al-2004a} Blain A.\ W., Chapman S.\ C., Smail I., Ivison R.\ J., 2004a, ApJ, 611, 52.
\bibitem[\protect\citeauthoryear{Blain}{2004b}]{Blain-et-al-2004b} Blain A.\ W., Chapman S.\ C., Smail I., Ivison R.\ J., 2004b, ApJ, 611, 725.
\bibitem[\protect\citeauthoryear{Borys}{2003}]{Borys-et-al-2003} Borys C., Chapman S.\ C., Halpern M., Scott D., 2003, MNRAS, 344, 385.
\bibitem[\protect\citeauthoryear{Braine \& Combes}{1992}]{Braine-and-Combes-1992} Braine J., Combes, F., 1992, A\&A, 264, 433.
\bibitem[\protect\citeauthoryear{Brown}{1991}]{Brown-and-Vanden-Bout-1991} Brown R.\ L., Vanden Bout P.\ A., 1991, AJ, 102, 1956.
\bibitem[\protect\citeauthoryear{Carilli}{2002a}]{Carilli-et-al-2002a} Carilli C.\ L.\ et al., 2002a, AJ, 123, 1838.
\bibitem[\protect\citeauthoryear{Carilli}{2002b}]{Carilli-et-al-2002b} Carilli C.\ L.\ et al., 2002b, ApJ, 575, 145.
\bibitem[\protect\citeauthoryear{Carilli}{2004}]{Carilli-et-al-2004} Carilli C.\ L.\ et al., 2004, ApJ, in press, (astro-ph/0409054).
\bibitem[\protect\citeauthoryear{Chapman}{2003a}]{Chapman-et-al-2003a} Chapman S.\ C.\ et al.\ 2003a, ApJ, 585, 57.
\bibitem[\protect\citeauthoryear{Chapman}{2003b}]{Chapman-et-al-2003b} Chapman S.\ C., Blain A.\ W., Ivison R.\ J., Smail I., 2003b, Nature, 422, 695. 
\bibitem[\protect\citeauthoryear{Chapman}{2003c}]{Chapman-et-al-2003c} Chapman S.\ C., Windhorst R., Odewahn S., Yan H., Conselice C.\ J., 2003c, ApJ, 599, 92.
\bibitem[\protect\citeauthoryear{Chapman}{2004}]{Chapman-et-al-2004} Chapman S.\ C., Smail I., Windhorst R., Muxlow T., Ivison R.\ J., 2004, ApJ, 611, 732.
\bibitem[\protect\citeauthoryear{Chapman}{2005}]{Chapman-et-al-2005} Chapman S.\ C., Blain A.\ W., Ivison R.\ J., Smail I., 2005, ApJ, in press, (astro-ph/0412573).
\bibitem[\protect\citeauthoryear{Cimatti}{1998}]{Cimatti-et-al-1998} Cimatti A., Andreani P., R\"{o}ttgering H., Tilanus R., 1998, Nature, 392, 895.
\bibitem[\protect\citeauthoryear{Cimatti}{2004}]{Cimatti-et-al-2004} Cimatti A.\ et al., 2004, Nature, 430, 184.
\bibitem[\protect\citeauthoryear{Cole}{1994}]{Cole-et-al-1994} Cole S., Aragon-Salamanca A., Frenk C.\ S., Navarro J.\ F., Zepf S.\ E., 1994, MNRAS, 271, 781.
\bibitem[\protect\citeauthoryear{Cole}{2000}]{Cole-et-al-2000} Cole S., Lacey C.\ G., Baugh C.\ M., Frenk C.\ S., 2000, MNRAS, 319, 168.
\bibitem[\protect\citeauthoryear{Cole}{2001}]{Cole-et-al-2001} Cole S.\ et al., 2001, MNRAS, 326, 255.
\bibitem[\protect\citeauthoryear{Condon}{1992}]{Condon-1992} Condon J.\ J., 1992, ARA\&A, 30, 575.
\bibitem[\protect\citeauthoryear{Cox}{2002}]{Cox-et-al-2002} Cox P.\ et al., 2002, A\&A, 387, 406.
\bibitem[\protect\citeauthoryear{Daddi}{2004}]{Daddi-et-al-2004} Daddi E., Cimatti A., Renzini A., 2004, ApJ, 600, L127.
\bibitem[\protect\citeauthoryear{De Breuck}{2003}]{De-Breuck-et-al-2003a} De Breuck C.\ et al., 2003a, A\&A, 401, 911.
\bibitem[\protect\citeauthoryear{De Breuck}{2003}]{De-Breuck-et-al-2003b} De Breuck C., Neri R., Omont A., 2003b, NewAR, 47, 285.
\bibitem[\protect\citeauthoryear{De Breuck}{2005}]{De-Breuck-et-al-2005} De Breuck C., Downes D., Neri R., van Breugel W., Reuland M., Omont A., Ivison R., 2005, A\&A, in press, (astro-ph/0411732).
\bibitem[\protect\citeauthoryear{Devereux}{1994}]{Devereux-et-al-1994} Devereux N., Taniguchi Y., Sanders D.\ B., Nakai N., Young J.\ S., 1994, AJ, 107, 2006.
\bibitem[\protect\citeauthoryear{Dey}{1999}]{Dey-et-al-1999} Dey A., Graham J.\  R., Ivison R.\ J., Smail I., Wright G.\ S., Liu M.\ C., 1999, ApJ, 519, 610.
\bibitem[\protect\citeauthoryear{Downes}{1995}]{Downes-et-al-1995} Downes D., Solomon P.\ M., Radford S.\ J.\ E., 1995, ApJ, 453, L65.
\bibitem[\protect\citeauthoryear{Downes \& Solomon}{1998}]{Downes-and-Solomon} Downes D., Solomon P.\ M., 1998, ApJ, 507, 615.
\bibitem[\protect\citeauthoryear{Downes}{1999}]{Downes-et-al-1999} Downes D., Neri R., Wiklind T., Wilner D.\ J., Shaver, P.\ A., 1999, ApJ, 513, L1.
\bibitem[\protect\citeauthoryear{Downes \& Solomon}{2003}]{Downes-and-Solomon-2003} Downes D., Solomon P.\ M., 2003, ApJ, 582, 37.
\bibitem[\protect\citeauthoryear{Drory}{2003}]{Drory-et-al-2003} Drory N., Bender R., Snigula J., Feulner, G., Hopp, U., Maraston C., Hill G.\ J., Mendes de Oliveira C., 2003, in The Masses of Galaxies at Low and High Redshift, ed. R. Bender \& A. Renzini (Berlin: Springer). (astro-ph/0201207). 
\bibitem[\protect\citeauthoryear{Dunne}{2000}]{Dunne-et-al-2000}  Dunne L., Eales S., Edmunds M., Ivison R.\ J., Alexander P., Clements D.\ L., 2000, MNRAS, 315, 115.
\bibitem[\protect\citeauthoryear{Fontana}{2004}]{Fontana-et-al-2004} Fontana A., Pozzetti L., Donnarumma I., 2004, A\&A, 424, 23.
\bibitem[\protect\citeauthoryear{Ford}{1994}]{Ford-et-al-1994} Ford H.\ C.\ et al., 1994, ApJ, 435, L27.
\bibitem[\protect\citeauthoryear{Frayer}{1998}]{Frayer-et-al-1998} Frayer D.\ T., Ivison R.\ J., Scoville N.\ Z., Yun M., Evans A.\ S., Smail I., Blain A.\ W., Kneib J.-P., 1998, ApJ, 506, L7.
\bibitem[\protect\citeauthoryear{Frayer}{1999}]{Frayer-et-al-1999} Frayer D.\ T.\ et al., 1999, ApJ, 514, 13L.
\bibitem[\protect\citeauthoryear{Frayer}{2003}]{Frayer-et-al-2003} Frayer D.\ T., Armus L., Scoville N.\ Z., Blain A.\ W., Reddy N.\ A., Ivison R.\ J., Smail Ian, 2003, ApJ, 126, 73.
\bibitem[\protect\citeauthoryear{Gao}{1999}]{Gao-et-al-1999} Gao Y., Solomon P.\ M., 1999, ApJ, 512, L99.
\bibitem[\protect\citeauthoryear{Gao \& Solomon}{2003}]{Gao-and-2003} Gao Y., Solomon P.\ M., 2003, ApJ, 606, 271.
\bibitem[\protect\citeauthoryear{Garrett}{2002}]{Garrett-2002} Garrett M.\ A., 2002, A\&A, 384, L19.
\bibitem[\protect\citeauthoryear{Genzel}{2003}]{Genzel-et-al-2003} Genzel R., Baker A.\ J., Tacconi L.\ J., Lutz D., Cox P., Guilloteau S., Omont A., 2003. ApJ, 584, 633.
\bibitem[\protect\citeauthoryear{Glazebrook}{2004}]{Glazebrook-et-al-2004} Glazebrook K.\ et al., 2004, Nature, 430, 181.
\bibitem[\protect\citeauthoryear{Granato}{2004}]{Granato-et-al-2004} Granato G.\ L., De Zotti G., Silva L., Bressan A., Danese, L., 2004, ApJ, 600, 580.
\bibitem[\protect\citeauthoryear{Greve}{2003}]{Greve-et-al-2003} Greve T.\ R., Ivison R.\ J., Papadopoulos P.\ P., 2003, ApJ, 599, 839.
\bibitem[\protect\citeauthoryear{Greve}{2004}]{Greve-et-al-2004a} Greve T.\ R., Ivison R.\ J., Papadopoulos P.\ P., 2004a, A\&A, 419, 99.
\bibitem[\protect\citeauthoryear{Greve}{2004}]{Greve-et-al-2004b} Greve T.\ R., Ivison R.\ J., Bertoldi F., Stevens J.\ A., Dunlop J.\ S., Lutz D., Carilli C.\ L., 2004b, MNRAS, 354, 779.
\bibitem[\protect\citeauthoryear{Guilloteau}{1997}]{Guilloteau-et-al-1997} Guilloteau S., Omont A., McMahon R.\ G., Cox P., Petitjean P., 1997, A\&A, 328, L1.
\bibitem[\protect\citeauthoryear{Guilloteau}{1999}]{Guilloteau-et-al-1999} Guilloteau S., Omont A., Cox P., McMahon R.\ G., Petitjean P., 1999, A\&A, 349, 363.
\bibitem[\protect\citeauthoryear{Guilloteau}{2000}]{Guilloteau-and-Lucas-2000} Guilloteau S., Lucas R., 2000, in "Imaging at Radio through Submillimeter Wavelengths", ed.\ J.\ G.\ Magnum \& S.\ J.\ E.\ Radford (San Francisco: ASP), 299. 
\bibitem[\protect\citeauthoryear{Hainline}{2004}]{Hainline-et-al-2004} Hainline L.\ J., Scoville N.\ Z., Yun M.\ S., Hawkins D.\ W., Frayer D.\ T., Isaak K.\ G., 2004, ApJ, 609, 61.
\bibitem[\protect\citeauthoryear{Holland}{1999}]{Holland-et-al-1999} Holland W.\ S.\ et al.\ 1999, MNRAS, 303, 659.
\bibitem[\protect\citeauthoryear{Hu \& Ridgway}{1994}]{Hu-and-Ridgway-1994} Hu E.\ M., Ridgway S.\ E., 1994, AJ, 107, 1303.
\bibitem[\protect\citeauthoryear{Iono}{2004}]{Iono-et-al-2004} Iono D., Yun M.\ S., Mihos J.\ C., 2004, ApJ, 616, 199.
\bibitem[\protect\citeauthoryear{Ivison}{2000}]{Ivison-et-al-2000} Ivison R.\ J.\ et al., 2000, MNRAS, 315, 209.
\bibitem[\protect\citeauthoryear{Ivison}{2001}]{Ivison-et-al-2001} Ivison R.\ J., Smail I., Frayer D.\ T., Kneib J.-P., Blain A.\ W., 2001, ApJ, 561, L45.
\bibitem[\protect\citeauthoryear{Ivison}{2002}]{Ivison-et-al-2002} Ivison R.\ J.\ et al., 2002, MNRAS,  337, 1.
\bibitem[\protect\citeauthoryear{Kauffmann}{1999}]{Kauffmann-et-al-1999} Kauffmann G., Colberg J.\ M., Diaferio A., White S.\ D.\ M., 1999, MNRAS, 303, 188.
\bibitem[\protect\citeauthoryear{Klamer}{2005}]{Klamer-et-al-2005} Klamer I.\ J., Ekers R.\ D., Sadler E.\ M., Weiss A., Hunstead R.\ W., De Breuck C., 2005, ApJL, in press, (astro-ph/0501447).
\bibitem[\protect\citeauthoryear{Kneib}{2004}]{Kneib-et-al-2004} Kneib J.-P., Neri R., Smail I., Blain A.\ W., Sheth K., van der Werf P., Knudsen K.\ K., 2004, A\&A, 614, L5.
\bibitem[\protect\citeauthoryear{Knudsen}{2003}]{Knudsen-et-al-2003} Knudsen K.\ K., van der Werf P.\ P., Jaffe W., 2003, A\&A, 411, 343.
\bibitem[\protect\citeauthoryear{Kooiman}{1995}]{Kooiman-et-al-1995} Kooiman B.\ L., Burns J.\ O., Klypin A.\ A., 1995, ApJ, 448, 500.
\bibitem[\protect\citeauthoryear{Kreysa}{1998}]{Kreysa-et-al-1998} Kreysa E.\ et al., 1998, SPIE, 3357, 319.
\bibitem[\protect\citeauthoryear{Krips}{2003}]{Krips-et-al-2003} Krips M., Neri R., Eckart A., Martin-Pintado J., Planesas P., Colina L., 2003, "Proceedings of the 4th  Cologne-Bonn-Zermatt-Symposium", ed. S. Pfalzner, C. Kramer, C. Straubmeier,  and A. Heithausen (Springer Verlag).
\bibitem[\protect\citeauthoryear{Ledlow}{2002}]{Ledlow-et-al-2002} Ledlow M.\ J., Smail Ian, Owen F.\ N., Keel W.\ C., Ivison R.\ J., Morrison G.\ E., 2002, ApJ, 577, L79.
\bibitem[\protect\citeauthoryear{Lewis}{2002}]{Lewis-et-al-2002} Lewis G.\ F., Carilli C.\ L., Papadopoulos P.\ P., Ivison R.\ J., 2002, MNRAS, 330, L15.
\bibitem[\protect\citeauthoryear{Lilly}{1998}]{Lilly-et-al-1998} Lilly S.\ et al., 1998, ApJ, 500, L75.
\bibitem[\protect\citeauthoryear{Martini \& Weinberg}{2001}]{Martini-and-Weinberg-2001} Martini P., Weinberg D.\ H., 2001, ApJ, 547, 12.
\bibitem[\protect\citeauthoryear{Mirabel \& Sanders}{1989}]{Mirabel-and-Sanders-1989} Mirabel I.\ F., Sanders D.\ B., 1989, 340, L53.
\bibitem[\protect\citeauthoryear{Neri}{2003}]{Neri-et-al-2003} Neri R.\ et al., 2003, ApJ, 597, L113.
\bibitem[\protect\citeauthoryear{Ohta}{1996}]{Ohta-et-al-1996} Ohta K., Yamada T., Nakanishi K., Kohno K., Akiyama M., Kawabe R., 1996, Nature, 382, 426.
\bibitem[\protect\citeauthoryear{Oke}{1995}]{Oke-et-al-1995} Oke J.\ B.\ et al., 1995, PASP, 107, 375.
\bibitem[\protect\citeauthoryear{Omont}{1996}]{Omont-et-al-1996} Omont A., Petitjean P., Guilloteau S., McMahon R.\ G., Solomon P.\ M., Pecontal E., 1996, Nature, 382, 428.
\bibitem[\protect\citeauthoryear{Omont}{2001}]{Omont-et-al-2001} Omont, A., Cox, P., Bertoldi, F., McMahon, R.\ G., Carilli, C., Isaak, K.\ G., 2001, A\&A, 374, 371.
\bibitem[\protect\citeauthoryear{Page}{2004}]{Page-et-al-2004} Page M., Stevens J.\ A., Ivison R.\ J., Carrera F.\ J., 2004, ApJ, 611, L85.
\bibitem[\protect\citeauthoryear{Papadopoulos}{2000}]{Papadopoulos-et-al-2000} Papadopoulos P.\ P., R\"{o}ttgering H.\ J.\ A., van der Werf P.\ P., Guilloteau S., Omont A., Breugel W.\ J.\ M., Tilanus R.\ P.\ J., 2000, ApJ, 528, 626.
\bibitem[\protect\citeauthoryear{Papadopoulos}{2001}]{Papadopoulos-et-al-2001} Papadopoulos P.\ P., Ivison R.\ J., Carilli C.\ L., Geraint L., 2001, Nature, 409, 58.
\bibitem[\protect\citeauthoryear{Papaopoulos \& Ivison}{2002}]{Papaopoulos-and-Ivison-2002} Papadopoulos P.\ P., Ivison R.\ J., 2002, ApJ, 564, L9.
\bibitem[\protect\citeauthoryear{Planesas}{1999}]{Planesas-et-al-1999} Planesas P., Martin-Pintado J., Neri R.,  Colina L., 1999, Science, 286, 2493.
\bibitem[\protect\citeauthoryear{Ravindranath}{2004}]{Ravindranath-et-al-2004} Ravindranath S.\ et al., 2004, ApJ, 604, L9.
\bibitem[\protect\citeauthoryear{Richards}{2002}]{Richards-et-al-2002} Richards G.\ T., Vanden Berk D.\ E., Reichard T.\ A., Hall P.\ B., Schneider D.\ P., SubbaRao M., Thakar A.\ R., York D.\ G., 2002, AJ, 124, 1.
\bibitem[\protect\citeauthoryear{Rickard}{1975}]{Rickard-et-al-1975} Rickard L.\ J., Palmer P., Morris M., Turner B.\ E., Zuckerman B., 1975, ApJ, 199, L75.
\bibitem[\protect\citeauthoryear{Rickard \& Harvey}{1984}]{Rickard-and-Harvey-1984} Rickard L.\ J., Harvey P.\ M., 1984, AJ, 89, 1520.
\bibitem[\protect\citeauthoryear{Rigopoulou}{1999}]{Rigopoulou-et-al-1999} Rigopoulou D., Spoon H.\ W.\ W., Genzel R., Lutz D., Moorwood A.\ F.\ M., Tran Q.\ D., 1999, AJ, 118, 2625.
\bibitem[\protect\citeauthoryear{Saracco}{2004}]{Saracco} Saracco P.\ et al., 2004, A\&A, 420, 125.
\bibitem[\protect\citeauthoryear{Sakamoto}{1999}]{Sakamoto-et-al-1999} Sakamoto K., Scoville N.\ Z., Yun M.\ S., Crosas M., Genzel R., Tacconi L.\ J., 1999, ApJ, 514, 68.
\bibitem[\protect\citeauthoryear{Sanders \& Mirabel}{1985}]{Sanders-and-Mirabel-1985} Sanders D.\ B., Mirabel I.\ F., 1985, ApJ, 298, L31.
\bibitem[\protect\citeauthoryear{Sanders}{1986}]{Sanders-et-al-1986} Sanders D.\ B., Scoville N.\ Z., Young J.\ S., Soifer B.\ T., Schloerb F.\ P., Rice W.\ L., Danielson G.\ E., 1986, ApJ, 305, L49.
\bibitem[\protect\citeauthoryear{Sanders}{1991}]{Sanders-et-al-1991} Sanders D.\ B., Scovillle N.\ Z., Soifer B.\ T., 1991, ApJ, 370, 158.
\bibitem[\protect\citeauthoryear{Scott}{2002}]{Scott-et-al-2002} Scott S.\ E.\ et al., 2002, MNRAS, 331, 817.
\bibitem[\protect\citeauthoryear{Scoville}{1997}]{Scoville-et-al-1997} Scoville N.\ Z., Yun M.\ S., Windhorst R.\ A., Keel W.\ C., Armus L., 1997, ApJ, 485, L21.
\bibitem[\protect\citeauthoryear{Seaquist}{1995}]{Seaquist-Ivison-Hall-1995} Seaquist E.\ R., Ivison R.\ J., Hall P.\ J., 1995, MNRAS, 276, 867.
\bibitem[\protect\citeauthoryear{Sheth}{2004}]{Sheth-et-al-2004} Sheth K., Blain A.\ W., Kneib J.-P., Frayer D.\ T., van der Werf P., Knudsen, K.\ K., 2004, ApJ, 614, L5.
\bibitem[\protect\citeauthoryear{Simpson}{2004}]{Simpson-et-al-2004} Simpson C., Dunlop J.\ S., Eales S.\ A., Ivison R.\ J., Scott S.\ E., Lilly S.\ J., Webb T.\ M.\ A., 2004, MNRAS, 353, 179.
\bibitem[\protect\citeauthoryear{Smail}{1997}]{Smail-et-al-1997} Smail I., Ivison R.\ J., Blain A.\ W., 1997, ApJ, 490, L5.
\bibitem[\protect\citeauthoryear{Smail}{1999}]{Smail-et-al-1999} Smail I., Ivison R.\ J., Kneib J.-P., Cowie L.\ L., Blain A.\ W., Barger A.\ J., Owen F.\ N., Morrison, G., 1999, MNRAS, 308, 1061.
\bibitem[\protect\citeauthoryear{Smail}{2002}]{Smail-et-al-2002} Smail I., Ivison R.\ J., Blain A.\ W., Kneib J.-P., 2002, MNRAS, 331, 495.
\bibitem[\protect\citeauthoryear{Smail}{2003}]{Smail-et-al-2003} Smail I., Chapman S.\ C., Ivison R.\ J., Blain A.\ W., Takata T., Heckman T.\ M., Dunlop J.\ S., Sekiguchi K., 2003, MNRAS, 342, 1185.
\bibitem[\protect\citeauthoryear{Smail}{2004}]{Smail-et-al-2004} Smail I., Chapman S.\ C., Blain A.\ W., Ivison R.\ J., 2004, ApJ, 616, 71.
\bibitem[\protect\citeauthoryear{Solomon \& Sage}{1988}]{Solomon-and-Sage-1998} Solomon P.\ M., Sage L.\ J., 1988, ApJ, 334, 613.
\bibitem[\protect\citeauthoryear{Solomon}{1992a}]{Solomon-et-al-1992a} Solomon P.\ M., Downes D., Radford S.\ J.\ E., 1992a, Nature, 356, 318.
\bibitem[\protect\citeauthoryear{Solomon}{1992b}]{Solomon-et-al-1992b} Solomon P.\ M., Downes D., Radford S.\ J.\ E., 1992b, ApJ, 398, L29.
\bibitem[\protect\citeauthoryear{Solomon}{1997}]{Solomon-et-al-1997} Solomon P.\ M., Downes D., Radford S.\ J.\ E., Barrett J.\ W., 1997, 478, 144. 
\bibitem[\protect\citeauthoryear{Solomon}{2003}]{Solomon-et-al-2003} Solomon P.\ M., Vanden Bout P., Carilli C.\ L., Guelin M., 2003, Nature, 426, 636.
\bibitem[\protect\citeauthoryear{Somerville}{2001}]{Somerville-et-al-2001} Somerville R.\ S., Primack J.\ R., Faber S.\ M., 2001, MNRAS, 320, 504.
\bibitem[\protect\citeauthoryear{Soucail}{1999}]{Soucail-et-al-1999} Soucail G., Kneib J.-P., Bezecourt J., Metcalfe L., Altieri B., Le Borgne J.\ F., 1999, A\&A, 343, L70.
\bibitem[\protect\citeauthoryear{Spergel}{2003}]{Spergel-et-al-2003} Spergel D.\ N.\ et al., 2003, ApJS, 148, 175.
\bibitem[\protect\citeauthoryear{Swinbank}{2004}]{Swinbank-et-al-2004} Swinbank A.\ M., Smail I., Chapman S.\ C., Blain A.\ W., Ivison R.\ J., Keel W.\ C., 2004, ApJ, 617, 64.
\bibitem[\protect\citeauthoryear{Tacconi}{2005}]{Tacconi-et-al-2005} Tacconi, L.\ et al., 2005, ApJ, submitted.
\bibitem[\protect\citeauthoryear{Tecza}{2004}]{Tecza-et-al-2004} Tecza M.\ et al., 2004, ApJ, 604, L109.
\bibitem[\protect\citeauthoryear{Walter}{2003}]{Walter-et-al-2003} Walter F.\ et al., 2003, Nature, 424, 406.
\bibitem[\protect\citeauthoryear{Webb}{2003}]{Webb-et-al-2003} Webb T.\ M.\ et al., 2003, ApJ, 587, 41.
\bibitem[\protect\citeauthoryear{Weiss}{2003}]{Weiss-et-al-2003} Weiss A., Henkel C., Downes D., Walter F., 2003, A\&A, 409, L41.
\bibitem[\protect\citeauthoryear{White}{1991}]{White-et-al-1991} White S.\ D.\ M., Frenk C.\ S., 1991, ApJ, 379, 52.
\bibitem[\protect\citeauthoryear{Wilner}{1995}]{Wilner-et-al-1995} Wilner D.\ J., Zhao, J.-H., Ho P.\ T.\ P., 1995, ApJ, 453, L91.
\bibitem[\protect\citeauthoryear{Wyse, Gilmore \& Franx}{1997}]{Wyse-et-al-1997} Wyse R.\ F.\ G., Gilmore G., Franx M., 1997, ARA\& A, 35, 637.
\bibitem[\protect\citeauthoryear{Yao}{2003}]{Yao-et-al-2003} Yao L., Seaquist E.\ R., Kuno N., Dunne L., 2003, ApJ, 588, 771. 
\bibitem[\protect\citeauthoryear{Young}{1984}]{Young-et-al-1984} Young J.\ S., Kenney J., Lord S.\ D., Schloerb F.\ P., 1984, ApJ, 287, L65. 
\bibitem[\protect\citeauthoryear{Young}{1986}]{Young-et-al-1986} Young J.\ S., Schloerb F.\ P., Kenney J.\ D., Lord S.\ D., 1986, ApJ, 304, 443. 
\bibitem[\protect\citeauthoryear{Yun}{2001}]{Yun-et-al-2001} Yun M.\ S., Reddy N.\ A., Condon  J.\ J., 2001, ApJ, 554, 803.
\bibitem[\protect\citeauthoryear{}{}]{}
\end{thebibliography}
\end{document}